\documentclass[runningheads]{llncs}
\usepackage{framed,multirow}

\usepackage{graphicx}

\usepackage{multirow, makecell}
\usepackage{xcolor}
\usepackage{graphicx}
\usepackage{enumitem}
\usepackage{ulem}

\usepackage{times}
\usepackage{epsfig}
\usepackage{graphicx}
\usepackage{amsmath}
\usepackage{amssymb}
\usepackage{wrapfig}

\usepackage{hyperref}
\usepackage{amssymb}
\usepackage{latexsym}

\usepackage{url}
\usepackage{xcolor}

\usepackage{hyperref}

\definecolor{newcolor}{rgb}{.8,.349,.1}

\usepackage{graphicx}

\usepackage{multirow, makecell}
\usepackage{xcolor}
\usepackage{graphicx}
\usepackage{enumitem}
\usepackage{ulem}

\usepackage{tikz}   
\usetikzlibrary{calc,arrows,positioning,shapes,shadows,spy,plotmarks,matrix,fit,backgrounds}

\usepackage{times}
\usepackage{epsfig}
\usepackage{graphicx}
\usepackage{amsmath}
\usepackage{amssymb}
\usepackage{wrapfig}
\begin{document}

\title{ASC-Net: Unsupervised Medical Anomaly Segmentation Using an Adversarial-based Selective Cutting Network}
\titlerunning{Raunak Dey et al}
%
\author{Raunak Dey \inst{1}, Wenbo Sun \inst{2}, Haibo Xu \inst{2} \and Yi Hong\inst{3}}
\institute{Department of Computer Science, University of Georgia\\ \email{raunak.dey@gmail.com} \and
Department of Radiology, Zhongnan Hospital of Wuhan University \and Department of Computer Science and Engineering, Shanghai Jiao Tong University \\ \email{yi.hong@sjtu.edu.cn}}

\maketitle 

\begin{abstract}
In this paper we consider the problem of unsupervised anomaly segmentation in medical images, which has attracted increasing attention in recent years due to the expensive pixel-level annotations from experts and the existence of a large amount of unannotated normal and abnormal image scans.
We introduce a segmentation network that utilizes adversarial learning to partition an image into two cuts, with one of them falling into a reference distribution provided by the user. This Adversarial-based Selective Cutting network (ASC-Net) bridges the two domains of cluster-based deep segmentation and adversarial-based anomaly/novelty detection algorithms. Our ASC-Net learns from normal and abnormal medical scans to segment anomalies in medical scans without any masks for supervision. We evaluate this unsupervised anomly segmentation model on three public datasets, i.e., BraTS 2019 for brain tumor segmentation, LiTS for liver lesion segmentation, and MS-SEG 2015 for brain lesion segmentation, and also on a private dataset for brain tumor segmentation. Compared to existing methods, our model demonstrates tremendous performance gains in unsupervised anomaly segmentation tasks. Although there is still room to further improve performance compared to supervised learning algorithms, the promising experimental results and interesting observations shed light on building an unsupervised learning algorithm for medical anomaly identification using user-defined knowledge.
\end{abstract}


\section{Introduction}
\label{section:intro}
 In the field of computer vision and medical image analysis, unsupervised image segmentation has been an active research topic for decades~\cite{giordana1997estimation,lee2002unsupervised,o2004combined,puzicha1999histogram,shi2000normalized}, due to its potential of applying to many applications without requiring the data to be manually labelled. Anomaly detection in images, which identifies out-of-distribution data samples or region of interests, often desires unsupervised algorithms, because of the complexity and large variation of anomalies and the challenge of obtaining a large amount of high-quality annotations for anomalies.
 
 Recently, advances in Generative Adversarial Networks (GANs)~\cite{goodfellow2014generative} have given rise to a class of anomaly detection algorithms, which are inspired by AnoGAN \cite{schlegl2017unsupervised} to identify abnormal events, behaviors, or regions in images or videos~\cite{del2016discriminative,erfani2016high,seebock2016identifying,naval2021implicit}. The AnoGAN approach learns a manifold of normal images by mapping the image space to a latent space based on GANs. To detect the anomaly in a query image, AnoGAN needs an iterative search in the latent space to reconstruct its closest corresponding image, which is subtracted from the query image for locating the anomaly. 
 That is, the AnoGAN family, including f-AnoGAN~\cite{schlegl2019f} and other following-up works~\cite{baur2018deep,berg2019unsupervised,kimura2018anomaly,zenati2018efficient,zenati2018adversarially}, focus on the reconstruction of the corresponding normal images for a query image, instead of directly working on the anomaly detection. As a result, their image reconstruction quality heavily affects the performance of anomaly detection or segmentation.

\begin{figure}[t]
\centering
\includegraphics[width=1\columnwidth]{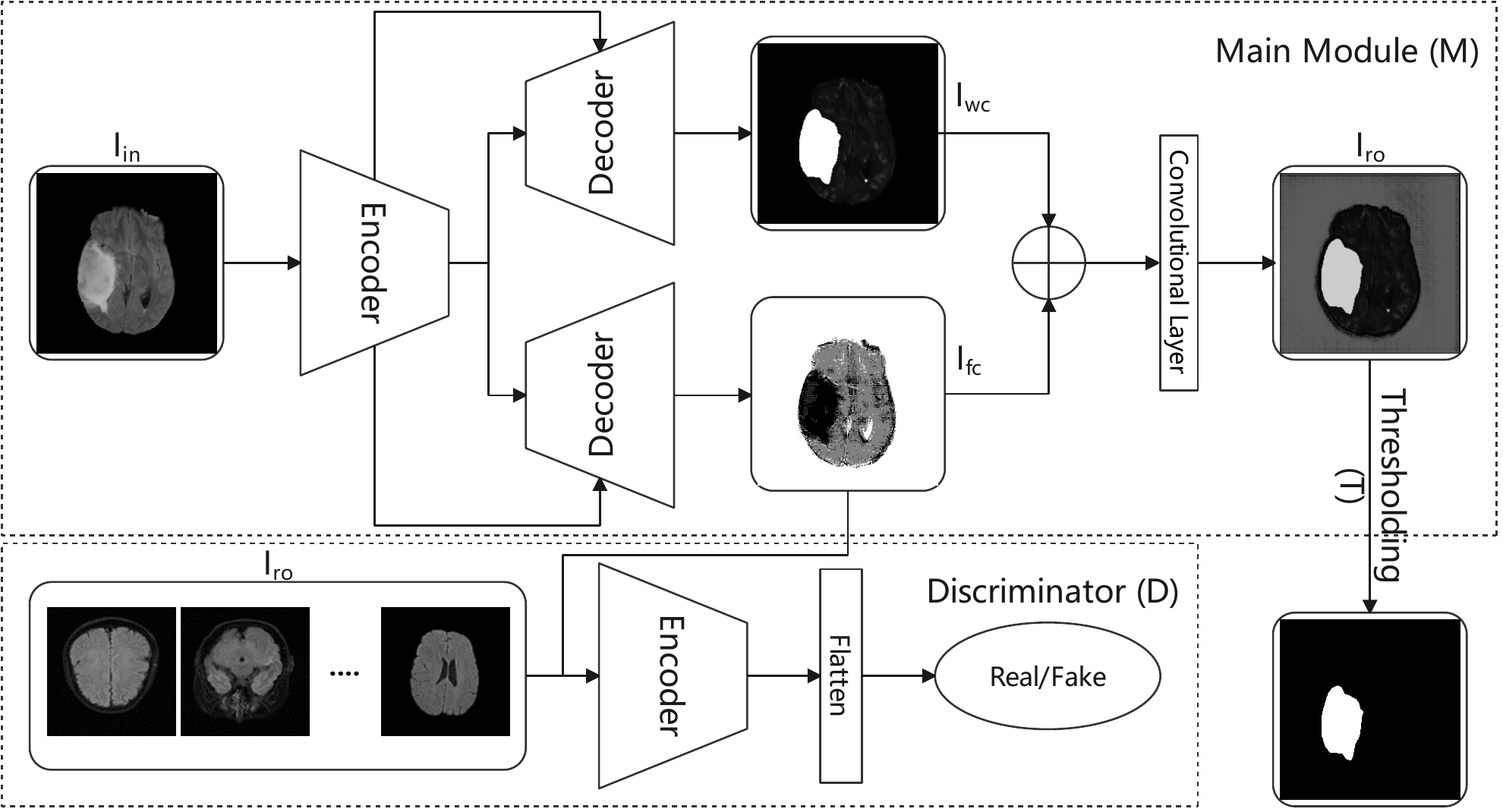}
\caption{Overview of our proposed adversarial-based selective cutting network (ASC-Net)  for unsupervised anomaly segmentation in medical scans.}
\label{fig:overview}
\end{figure}

To center the focus on the anomaly without needing faithful reconstruction, we propose an adversarial-based selective cutting neural network (ASC-Net), as shown in Figure~\ref{fig:overview}. Our goal is to decompose an image into two selective cuts and simplify the task of anomaly segmentation, according to a collection of normal images as a reference distribution. Typically, this reference distribution is defined by a set of images provided by users or experts who have vague knowledge and expectation of normal control cases. In this way, one cut of an image will fall into the reference distribution, while other image content that is outside of the reference image distribution will be grouped into the other cut. These two cuts allow to reconstruct the original input image \textit{semantically}, resulting in a simplified reconstruction where clustering normal and abnormal regions could be easily achieved by performing an intensity thresholding. 

In the architecture design shown in Figure~\ref{fig:overview}, we consider the above two cuts simultaneously in the U-Net framework~\cite{ronneberger2015u}, which is extended with two upsampling branches and one connecting to a GAN's discriminator network. This discriminator allows introducing the knowledge contained in the reference image distribution and makes the main network decompose images into softly disjoint two regions. That is, the generation of our selective cuts is under the constraint of the reference image distribution. We keep the reconstruction after the two-cut generation simple to preserve the semantic information in the two cuts and make the anomaly segmentation possible based on thresholding. As a result, we obtain a joint estimation of anomaly segmentation and a reduced corresponding normal image, thus bypassing the need for a perfect reconstruction. Except for a collection of normal images in the reference distribution, we do not have any other labels at the image level or annotations at the pixel/voxel level for the anomaly; hence, our method becomes an unsupervised solution for anomaly detection and segmentation in medical images.

We evaluate our proposed unsupervised anomaly segmentation network on three public datasets, i.e., MS-SEG2015~\cite{carass2017longitudinal}, BraTS-2019~\cite{bakas2017advancing,bakas2018identifying,menze2014multimodal}, and LiTS~\cite{bilic2019liver} datasets. For the MS-SEG2015 dataset, an exhaustive study on comparing multiple existing autoencoder-based models, variational-autoencoder-based models, and GAN-based models is performed in~\cite{baur2021autoencoders}.
Compared to the best Dice scores reported in~\cite{baur2021autoencoders}, we have significant gains in performance, with an improvement of 23.24\% in mean Dice score without any post-processing and 20.40\% with post-processing\footnote{Different from that in~\cite{baur2021autoencoders}, we use a simple open-and-closed operation as the post-processing.}.
For BraTS dataset, our experiments show that f-AnoGAN, the one performs the best after post-processing in~\cite{baur2021autoencoders}, has difficulty reconstructing the normal images required for anomaly segmentation. By contrast, under the two-fold cross-validation settings we obtain a mean Dice score of 63.67\% for the BraTS brain tumor segmentation before post-processing and 68.01\% after post-processing. Similarly, for the LiTS liver lesion segmentation, we obtain 32.24\% before post-processing and improve the mean Dice score to 50.23\% using a simple post-processing scheme of open and closed sets. We also evaluate our method on a private dataset to further demonstrate its effectiveness and generalization in practice, which achieves over 90\% mean Dice score after post-processing.

\vspace{0.05in}
\noindent 
Overall, the contributions of our proposed method in this paper are summarized below:
\begin{itemize}[noitemsep,topsep=0pt]
\itemsep0em 
    \item Proposing an adversarial based framework for unsupervised anomaly segmentation, which bypasses the normal image reconstruction and works on anomaly segmentation directly. This framework presents a general strategy to generate two selective cuts with incorporating human knowledge via a reference image set. 
    \item To the best of our knowledge,  when we proposed model ours was the first one to apply an unsupervised segmentation algorithm to both BraTS 2019 and LiTS liver lesion public datasets. Also, our method outperforms the AnoGAN family and other popular methods presented in~\cite{baur2021autoencoders} on the publicly available MS-SEG2015 dataset.
\end{itemize}

This paper is an extension of our MICCAI paper published in 2021~\cite{dey2021ascnet}. To further demonstrate the effectiveness of our proposed method, we provide more experimental results, including comparisons to a new algorithm that becomes available online after the MICCAI conference and additional evaluations on a private dataset. In addition, compared to the conference version, this paper presents more detailed descriptions and explanations on motivations and design choices, new observations while working on the private dataset, and thorough discussions on the merits and limitations of our method. 

\section{Related Work}
\label{related_ch6}

In this paper, we perform unsupervised segmentation for anomalies in medical images via clustering. Therefore, we review the related work from these two perspectives, i.e., clustering and anomaly segmentation in an unsupervised manner. 

\subsection{Unsupervised Segmentation Based on Clustering}
Before the prevalence of deep neural networks, the primary means of medical image segmentation involve a combination of classical clustering operations followed by edge detection, watershed transformation, k-means, such as in~\cite{grau2004improved,ng2006medical,1633722}. Since deep learning technique becomes popular in 2012, it opens a new venue to develop more efficient and effective image segmentation methods.   W-Net~\cite{xia2017w} is a recent representative unsupervised image segmentation approach, which introduces the classical graph clustering method~\cite{shi2000normalized} into deep neural networks to find k-clusters in an image. Since W-Net requires the computation of the similarity score matrix for all pixels, resulting in its high-computational cost for high resolution images. Moreover, W-Net has difficulty in generating consistent cut outputs due to the uncertainty in the network initialization.




\subsection{Unsupervised Anomaly Detection and Segmentation}

The primary school of thought in the case of unsupervised anomaly detection and segmentation has revolved around the idea of training frameworks to learn the manifold of healthy samples, which helps filter out anomalies in a query image during testing for reconstruction and comparison. GAN-based and AutoEncoder (AE) based algorithms are the two most representative categories in addressing the unsupervised anomaly detection and segmentation problem. 



\noindent
\textbf{GAN-Based Methods.}
The majority of GAN-based applications~\cite{berg2019unsupervised,kimura2018anomaly,zenati2018efficient,zenati2018adversarially} are motivated from the pioneering paper AnoGAN~\cite{schlegl2017unsupervised}, which is later improved with a faster version f-AnoGANs~\cite{schlegl2019f}. The principle idea of all methods in the AnoGAN's family involves training a neural network in an adversarial fashion to learn and generate perfect healthy images. In this way, any unhealthy or anomalous image would not be constructive faithfully, and thus a residual or subtraction between the regenerated image during testing and its output image would provide the anomaly.

\noindent
\textbf{AE-Based Methods.}
Following the success of AnoGAN idea of anomaly detection and accompanied by advances in Variational Autoencoders (VAE)~\cite{kingma2013auto,kingma2014stochastic}, adversarial autoencoders in~\cite{makhzani2015adversarial} replace the KL-divergence loss in VAE with an adversarial network to ensure the generation of meaningful samples from prior distribution. Similar combination has also been proposed in~\cite{pmlr-v48-larsen16} to create a merger between GANs and VAEs, which has been utilized for anomaly segmentation in brain MRI scans~\cite{baur2018deep}. Different variations of VAEs have also been employed in~\cite{chen2018deep,chen2020unsupervised,zimmerer2018context} for brain tumor detection.


\subsection{Comparison with Existing Methods}
\label{motivation_ch6}
One main difference between existing methods and ours is the role that image reconstruction plays in anomaly detection. AnoGANs and approaches inspired by AnoGAN \cite{baur2018deep,berg2019unsupervised,kimura2018anomaly,zenati2018adversarially}, as well as the VAE variants, focus on regenerating the corresponding normal images for a query image, but not directly working on the anomaly detection. The performance of these methods on anomaly detection highly depends on the image reconstruction quality, which makes the methods susceptible to GANs instability issue during reconstruction. Failing to reconstruct a corresponding normal image will lead to inaccurate prediction results for the anomaly detection and segmentation task. On the contrary, we aim to develop a solution that is constrained by the underlying normal images and generates a \textit{reduced} reconstruction of the input image for segmentation based on a simple thresholding. Our method does generate a healthy component of an input image; however, its generation quality is not used as a termination criteria for our algorithm. Instead, a clear histogram separation with distinct peaks of the reduced reconstruction is the requirement for the termination. This design choice frees our framework from the instability of the adversarial training scheme.

Another main difference is the way to define disjoint groups of an image. 
Unsupervised segmentation and anomaly detection both aim to produce clusters that separate image pixels into disjoint groups. In the anomaly detection algorithms, the AnoGANs family~\cite{schlegl2019f,schlegl2017unsupervised} separates a query image into the normal and abnormal groups by generating high-quality normal images based on GANs, while graph-based methods like W-Net~\cite{xia2017w} could perform two cuts to obtain two clusters, i.e., normal and anomalous. Our approach, on the contrary, defines the disjoint groups at semantic level of the input image, based on the complement laws in set theory. In this way, we decompose the input image into two sets, according to the group of pixels in the sets, not simply based on the pixel intensity itself. As a result, we achieve better/robust results compared to methods in AnoGAN's family and W-Net\footnote{We tried W-Net, but it cannot provide consistent outputs at each run.}. 


A recent work~\cite{zhang2021self} introduces a self-learning approach, which creates an artificial dataset to pretrain a segmentation network and test on similar datasets. That is, for different segmentation tasks, this approach needs to create a similar artificial dataset with Pseduo annotations, which has the risk of introducing label noise and segmentation artifacts. Unlike this method, our framework is general and does not produce artificial datasets for different tasks. For example, we are able to segment brain Lesions and brain tumors using non-tumor slices collected from the BraTS dataset, as shown in the experiments.

\section{Adversarial-based Selective Cutting Network (ASC-Net)}

\subsection{Network Framework}
\label{sec:framework}

Our network aims to decompose an input image into two discontinuous sub-images, which is achieved by the main module $M$ shown in Fig.~\ref{fig:overview}. To guide the decomposition, we incorporate the user knowledge defined by a reference distribution of normal images, using the discriminator $D$. These two modules are the core components of our proposed ASC-Net, which is followed by a simple clustering step (the thresholding $T$) to obtain the segmentation mask of the anomaly. 

\begin{figure*}[t]
\centering
  \includegraphics[height=0.115\textwidth]{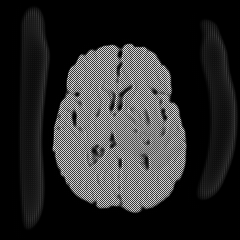}
  \includegraphics[height=0.115\textwidth,width=0.2\textwidth]{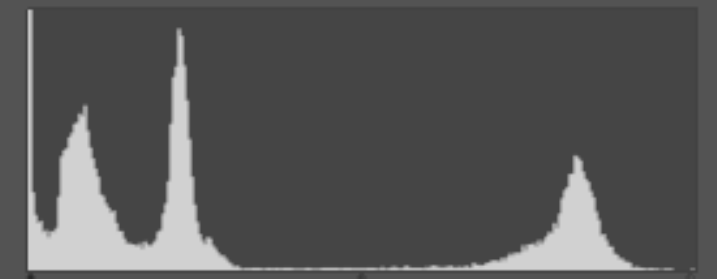}
  \includegraphics[height=0.115\textwidth]{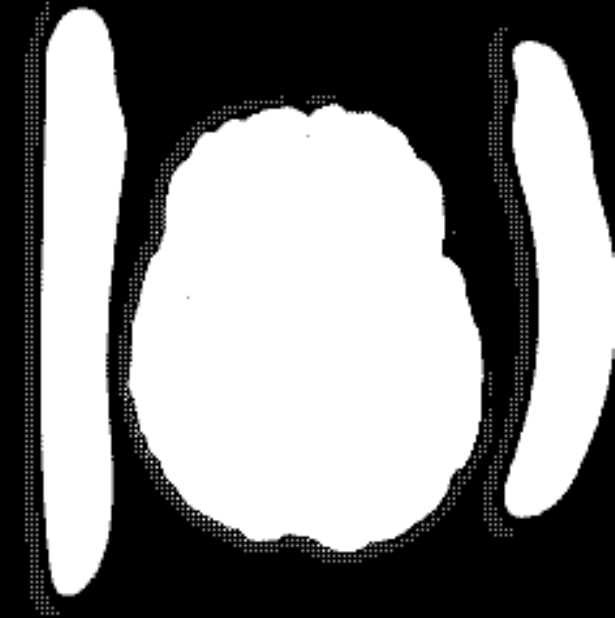}
  \includegraphics[height=0.115\textwidth]{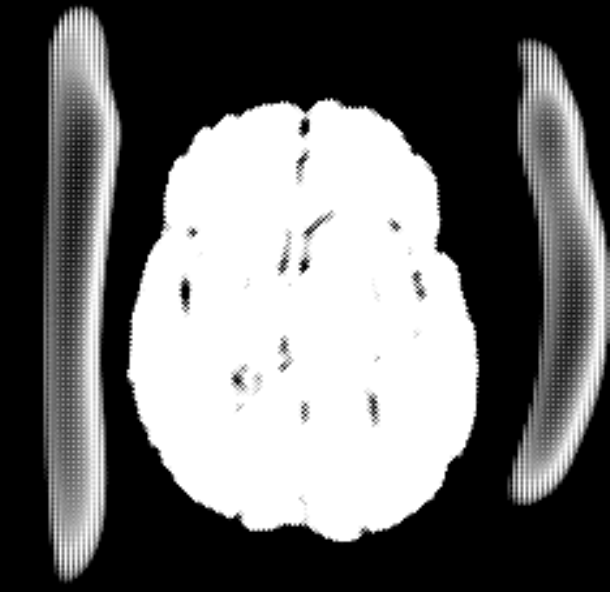}
  \includegraphics[height=0.115\textwidth]{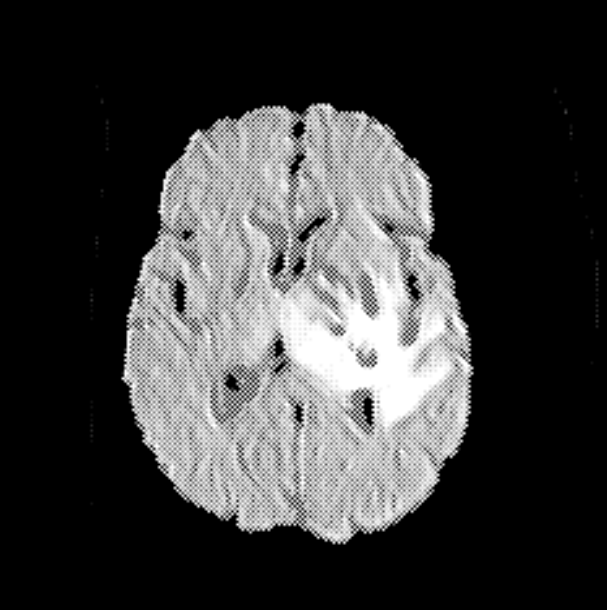}
  \includegraphics[height=0.115\textwidth]{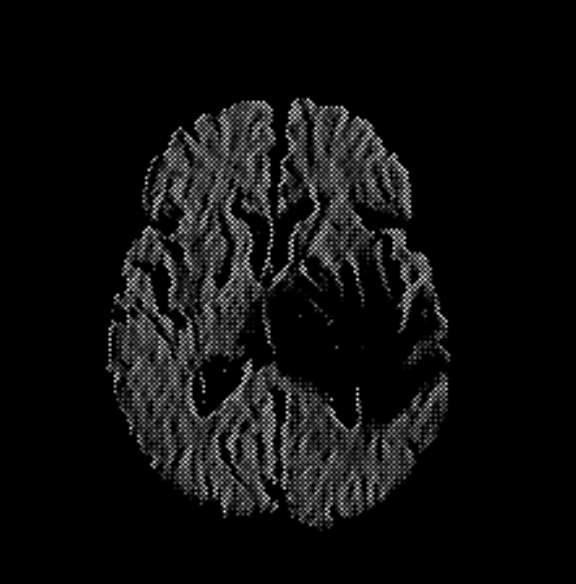}\\
  \includegraphics[height=0.116\textwidth]{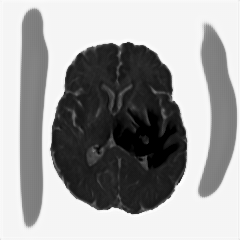}
  \includegraphics[height=0.116\textwidth,width=0.2\textwidth]{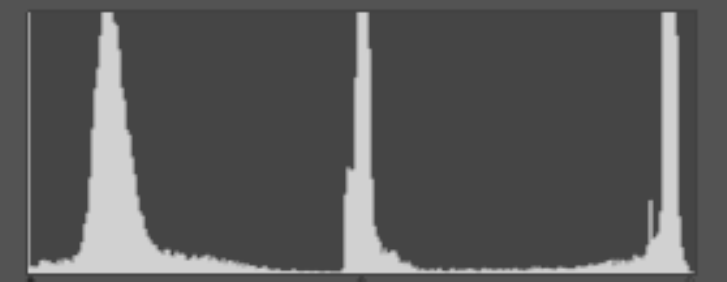}
  \includegraphics[height=0.116\textwidth]{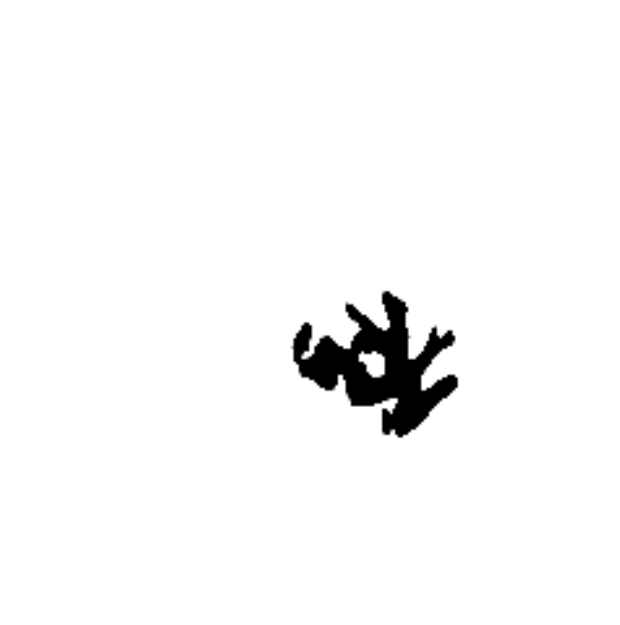}
  \includegraphics[height=0.116\textwidth]{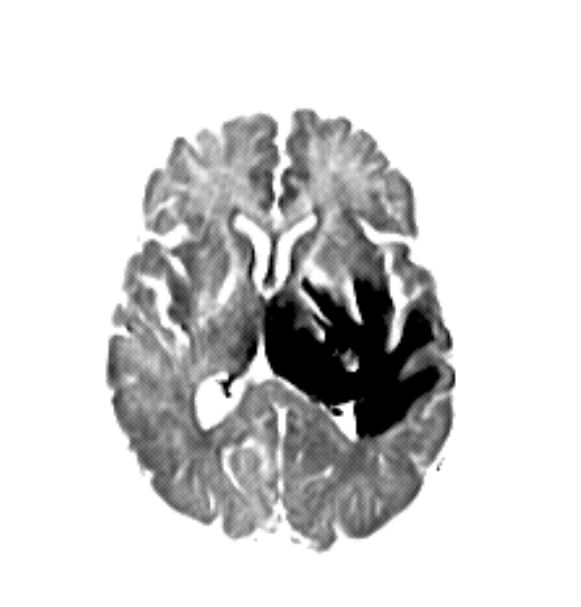}
  \includegraphics[height=0.116\textwidth]{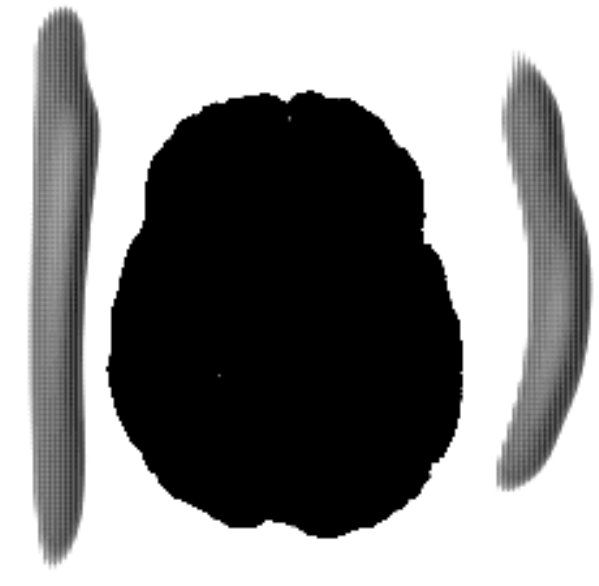}
  \includegraphics[height=0.116\textwidth]{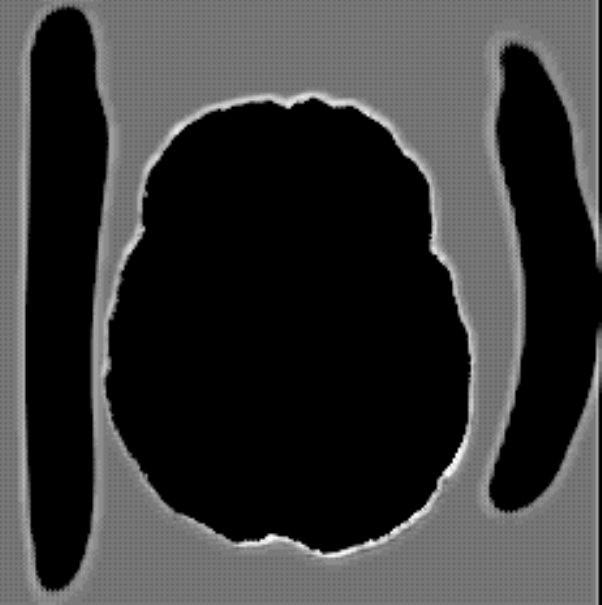}
\caption{Visualization of the “disjoincy” between images $I_{fc}$ (top) and $I_{wc}$ (bottom) generated by two cuts of ASC-Net. Left to right: the generated image, its histogram, and the following four columns representing the histogram equalized images of the thresholded peaks with the first peak being the first image, etc. The first peak of $I_{fc}$ is disjoint with the last peak of $I_{wc}$, etc.}
\label{fig:proof_dis}
\end{figure*}

\vspace{0.05in}
\noindent
\textbf{Main Module $M$.} The main module aims to generate two selective cuts, which separate the normal and abnormal information in an input image. 
The $M$ follows an encoder-decoder architecture like the U-Net, including one encoder and two decoders. The encoder $E$ extracts features of an input image $I_{in}$, which could be an image located within or outside of the reference distribution $\{I_{rd}\}$, a collection of normal images.
One decoder, the second upsampling branch, is designed to generate a ``fence'' cut $C_f$ that is constrained by an image fence formed by $\{I_{rd}\}$. The $C_f$ aims to generate an image $I_{fc}$ and tries to fool the discriminator $D$. The other decoder, the first unsampling branch, is designed to generate another ``wild'' cut $C_w$, which captures leftover image content that is not included in $I_{fc}$. As a result, the $C_w$ produces another images $I_{wc}$ to complement the fence-cut output $I_{fc}$. Since the wile-cut output is complementary to the fence-cut output, image information that can not be covered by the reference distribution would be included in the while-cut output, like the anomaly. The complementary relation between these two cuts $C_f$ and $C_w$ is enforced by a positive Dice loss, i.e., a ``disjoincy" loss as discussed in the paragraph of Loss Functions. Figure~\ref{fig:proof_dis} demonstrates the ``disjoincy" of $I_{fc}$ and $I_{wc}$, like their complementary histogram distribution and different thresholded images at different peaks.



\begin{figure}[h]
\centering
  \includegraphics[height=0.115\textwidth,width=0.115\textwidth]{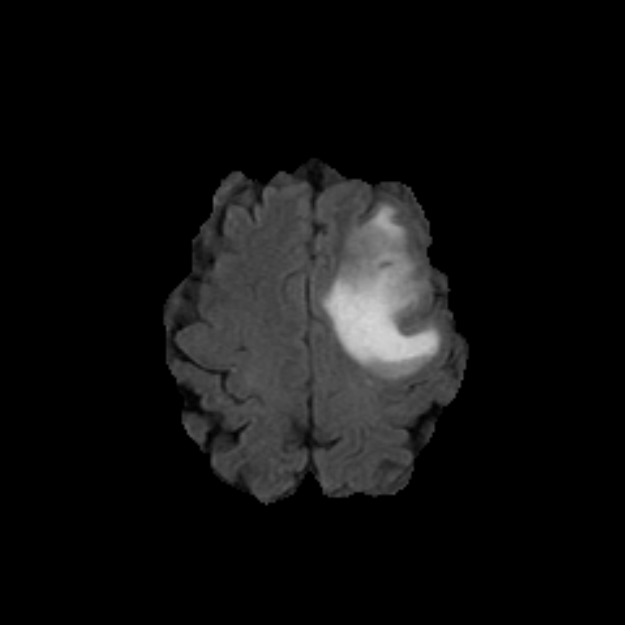}
  \includegraphics[height=0.115\textwidth,width=0.115\textwidth]{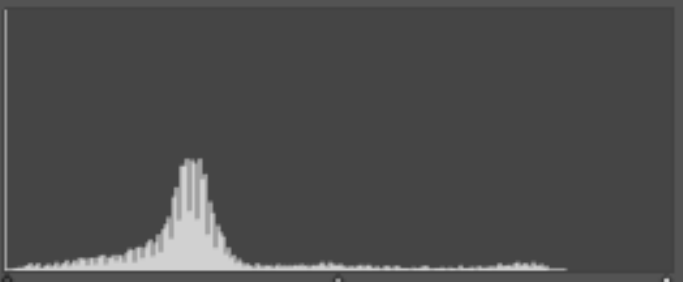}
  \includegraphics[height=0.115\textwidth,width=0.115\textwidth]{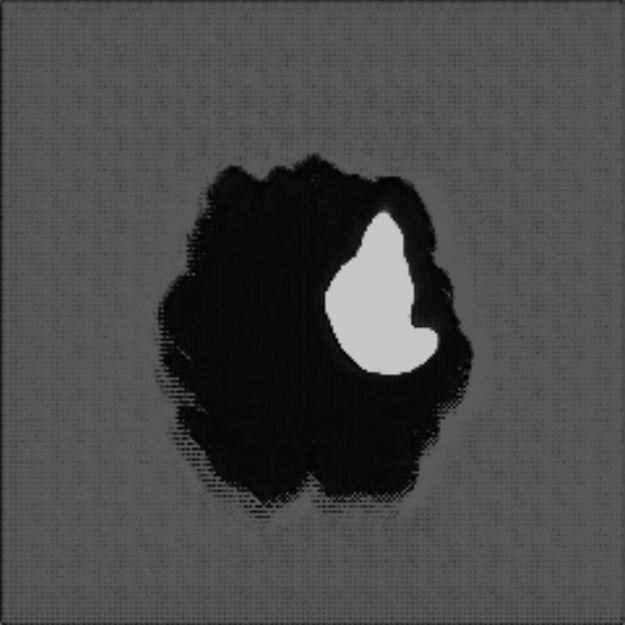}
  \includegraphics[height=0.115\textwidth,width=0.115\textwidth]{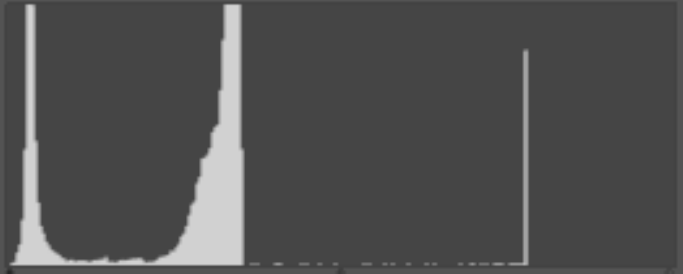}\\
    \includegraphics[height=0.115\textwidth,width=0.115\textwidth]{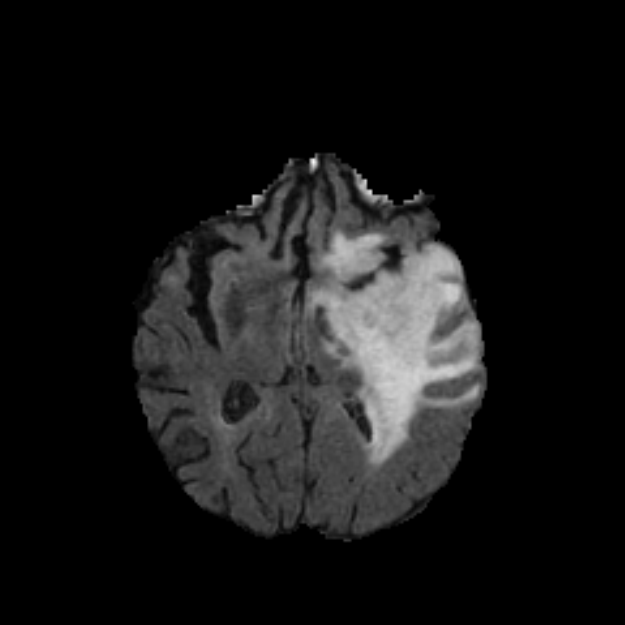}
  \includegraphics[height=0.115\textwidth,width=0.115\textwidth]{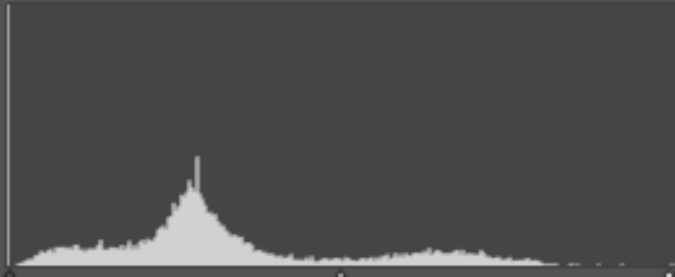}
  \includegraphics[height=0.115\textwidth,width=0.115\textwidth]{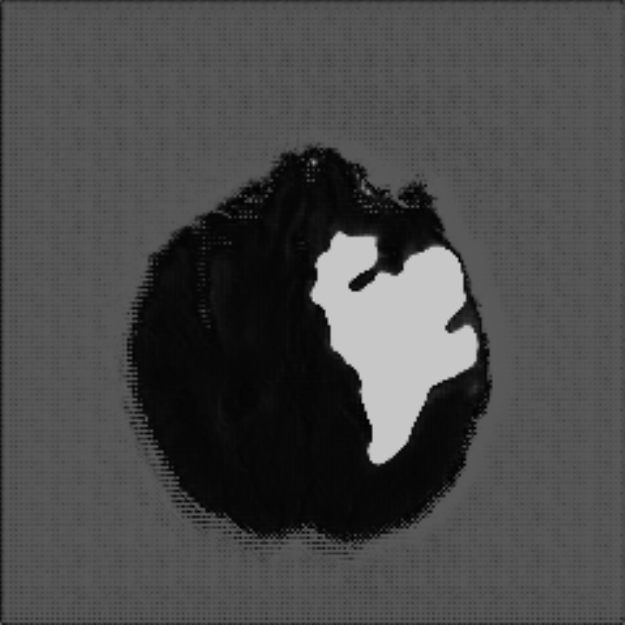}
  \includegraphics[height=0.115\textwidth,width=0.115\textwidth]{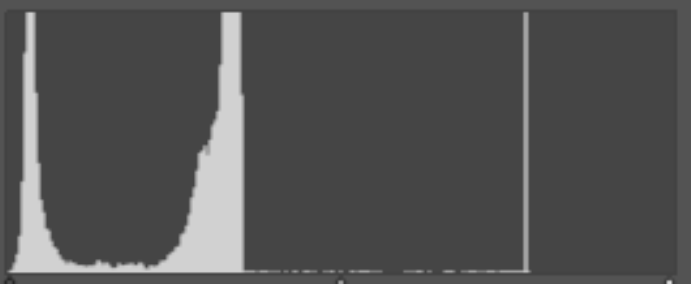}
\caption{Histogram comparison of two sample images. Left to right: the input image, its histogram, its reconstructed image using ASC-Net, and the histogram of the reconstructed image. The histograms of the input images vary greatly, while the ones of their reconstructions show peaks at similar ranges, which enables a thresholding-based pixel-level separation.}
\label{fig:input_vs_op}
\end{figure}

Except for a weak guidance of the ``disjoincy" loss for the ``wild" cut, we adopt a reconstruction branch to make sure our two selective cuts include enough information to constrcut a coarse version of the original input.
The reconstructor $R$ consists of a $1 \times 1$ convolution layer with the Sigmoid as the activation function, which is applied on the concatenation of the two-cut outputs $I_{fc}$ and $I_{wc}$ to regenerate the input image $I_{in}$ back. This reconstructor $R$ ensures that the $C_f$ does not generate an image $I_{fc}$ far from the input image $I_{in}$ and also ensures that the $C_w$ does not generate an empty image $I_{wc}$ if the anomaly or novelty exists. Figure~\ref{fig:input_vs_op} shows the histogram separation of the reconstructed images, compared to the original input images which present complex histogram peaks and have difficulty in separating the brain tumor from backgorund and other tissues via a simple thresholding. The discontinuous histogram distribution of $I_{ro}$ is inherited from the two generated sub-images $I_{fc}$ and $I_{wc}$ through a simple weighted combination. As a result, the segmentation task becomes relatively easy to be done on the reconstructed image $I_{ro}$.

\vspace{0.05in}
\noindent
\textbf{Discriminator $D$.} The GAN discriminator tries to distinguish the generated image $I_{fc}$, according to a reference distribution $R_d$ defined by a set of images $\{I_{rd}\}$, which are provided by the user or experts. The $R_d$ typically includes images collected from the same group, for instance, normal brain scans, which share similar structures and lie on a manifold. Introducing $D$ allows us to incorporate our vague prior knowledge about a task into a deep neural network. Typically, it is non-trivial to explicitly formulate such prior knowledge; however, it could be implicitly represented by a selected image set. The $R_d$ is an essential component that makes our ASC-Net possible to generate selective cuts according to the user's input, without requiring other supervisions or pixel-level annotations. 

\vspace{0.05in}
\noindent
\textbf{Thresholding $T$.}
The reconstruction branch consistently provides us a reduced version of the original input image, where an unsupervised segmentation is easy to achieve via clustering.  
To cluster the reconstructed image $I_{ro}$ into two groups at the pixel level, we choose the thresholding approach with the threshold values obtained using the histogram of $I_{ro}$. We observed that for an anomaly that is often brighter than the surrounding tissues like the BraTS brain tumor, the intensity value at the rightmost peak of the histogram is a desired threshold; while an opposite case like darker LiTS liver lesions, the value at the leftmost peak would be the threshold. We also observed that the histograms of the reconstructed images for different inputs reflect the same cut-off point for the left or right peaks, which allows using one threshold for an entire dataset.

\vspace{0.05in}
\noindent
\textbf{Loss Functions.}
The main module $M$ includes three loss functions: (i) the image generation loss for $C_f$, $Loss_{C_f}$, (ii) the ``disjoincy" loss between $C_f$ and $C_w$, $Loss_{C_w}$, and (iii) the reconstruction loss, $Loss_{R}$.   
In particular, the $C_f$ tries to generate an image $I_{fc}$ that fools the discriminator $D$ by minimizing $Loss_{C_f} = \frac{1}{n}\sum_{i=1}^{n} |D(I_{fc}^{(i)})-1|$. Here, $n$ is the number of samples in the training batch. The $C_w$ tries to generate an image $I_{wc}$ that is complement to $I_{fc}$ by minimizing the positive Dice score $Loss_{C_w} = \frac{2 |I_{fc} \cap I_{wc}|}{|I_{fc}| + |I_{wc}|}$.
The last reconstruction takes an Mean-Squared-Error (MSE) loss between the input image $I_{in}$ and the reconstructed image $I_{ro}$:
    $Loss_R = \frac{1}{n}\sum_{i=1}^{n}\|I_{in}^{(i)} - I_{ro}^{(i)}\|_2^{2}$. 
The discriminator $D$ tries to reject the the $C_f$ output $I_{fc}$ but accept the images from the reference distribution $R_d$, by minimizing the following loss function:
    $Loss_D = \frac{1}{n+m} \left(\sum_{i=1}^{n} |D(I_{fc}^{(i)})-(-1)| +\sum_{i=1}^{m} |D(I_{R_d}^{(i)})-1| \right)$.
Here, $m$ is the number of the images in $R_d$. Even though $D$ and $C_{fc}$ are tied in an adversarial setup, here we do not use the Earth Mover distance~\cite{rubner2000earth} in the loss function, since we would like $D$ to identify both positive samples and negative samples with equal precision. Therefore, we use Mean Absolute Error (MAE) instead. 


\subsection{Architecture Details and Training Scheme}
\label{sec:training}
We use the same network architecture for all of our experiments as shown in Fig.~\ref{fig:overview}. 
The encoder $E$ consists of four blocks of two convolution layers with a filter size of $(3,3)$ followed by a max pooling layer with a filter size of $(2,2)$ and batch normalization after every convolution layer. After every pooling layer we also introduce a dropout of 0.3. The number of feature maps in each of the convolution layer of a block are 32, 64, 128, and 256. Following these blocks is a transition layer of two convolution layers with feature maps of size 512 followed by batch normalization layers. 
The $C_{fc}$ and $C_{wc}$ decoders are connected to the $E$ and mirror the layers with the pooling layers replaced with 2D transposed convolutional layers, which have the same number of feature maps as the blocks mirror those in the encoder. Similar to a U-Net, we also introduce skip connections across similar levels in the encoder and decoders. The reconstructor $R$ is simply a Sigmoid layer applied to the concatenation of $I_{fc}$ and $I_{wc}$, resulting in a simplified CompNet~\cite{dey2018compnet}. 
The Discriminator $D$ mimics the architecture of the $E$, except for the last layer where a dense layer is used for classification. All the intermediate layers have ReLU activation function and the final output layers have the Sigmoid activation. The only exception is the output of the discriminator $D$, which has a Tanh activation function to separate $I_{fc}$ and images from the $R_d$ to the maximum extent. 

\vspace{0.05in}
\noindent
We use Keras with Tensorflow backend and Adam optimizer with a learning rate of 5e-5 to implement our architecture. 
We follow two distinct training stages:
\begin{itemize}[noitemsep, nolistsep]
\item In the {\it first stage}, we train $D$ and $M$ in cycles. We start training $D$ with $\{R_d\}$ with True labels and $\{I_{fc}\}$ with False labels. These training samples are shuffled randomly. Following $D$, we train $M$ with $\{I_{in}\}$ as input and the weights of $D$ frozen while preserving the connection between $\{I_{fc}\}$ and $D$. The objective of the $M$ is to morph the appearance of $\{I_{in}\}$ into $\{I_{fc}\}$ to fool $D$ with the frozen weights. We call these two steps one cycle, and in each step there may be more than one epochs of training for $M$ or $D$. 



\item In the {\it second stage}, $M$ and $D$ continue to be trained alternatively; however, the input images to $D$ are changed, since the training purpose at this stage is to focus on the differences between the $\{R_d\}$ and $\{I_{in}\}$, while ignoring the noisy biases created by the $M$ in transforming $\{I_{in}\}$ to $\{I_{fc}\}$. To achieve this, we augment the reference distribution $\{R_d\}$ with its generated images via $M$, i.e., $\{I_{fc}(R_d)\}$. We treat them as true images, and the union set $\{R_d \cup I_{fc}(R_d)\}$ is used to update $D$. 

\end{itemize}

\begin{figure*}[t]
\centering
\setlength\tabcolsep{1.5pt}
\begin{tabular}{ccccccc}
$I_{in}$ & $I_{fc}$ & $I_{wc}$ & $I_{ro}$ & $M_{gt}$& $M_{pred}$ & $M_{pred} \cap I_{in}$ \\

\includegraphics[width=0.13\textwidth]{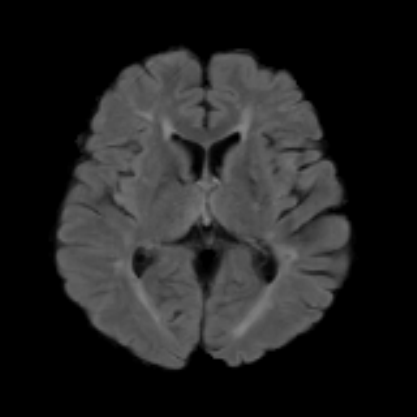} &
\includegraphics[width=0.13\textwidth]{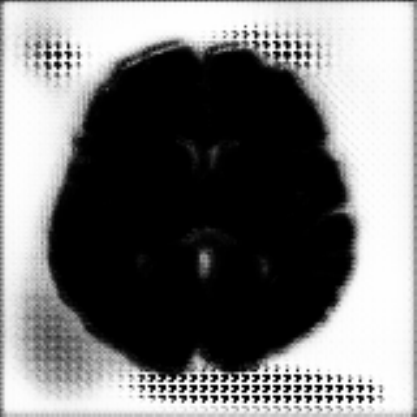} &
\includegraphics[width=0.13\textwidth]{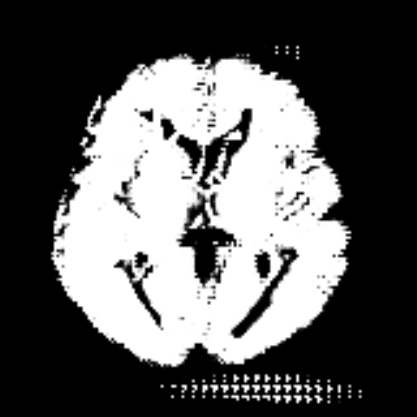} &
\includegraphics[width=0.13\textwidth]{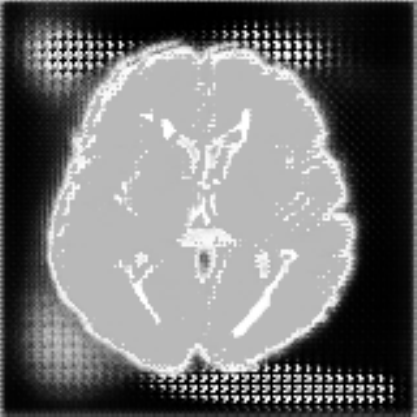} &
\includegraphics[width=0.13\textwidth]{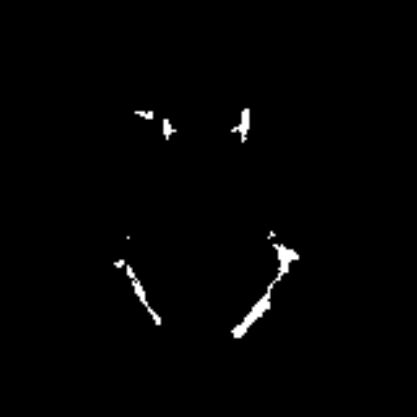} &
\includegraphics[width=0.13\textwidth]{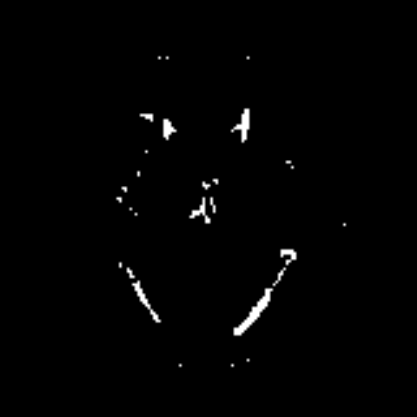} &
\includegraphics[width=0.13\textwidth]{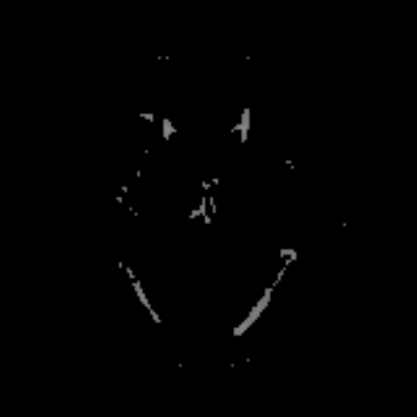} \\

\includegraphics[width=0.13\textwidth]{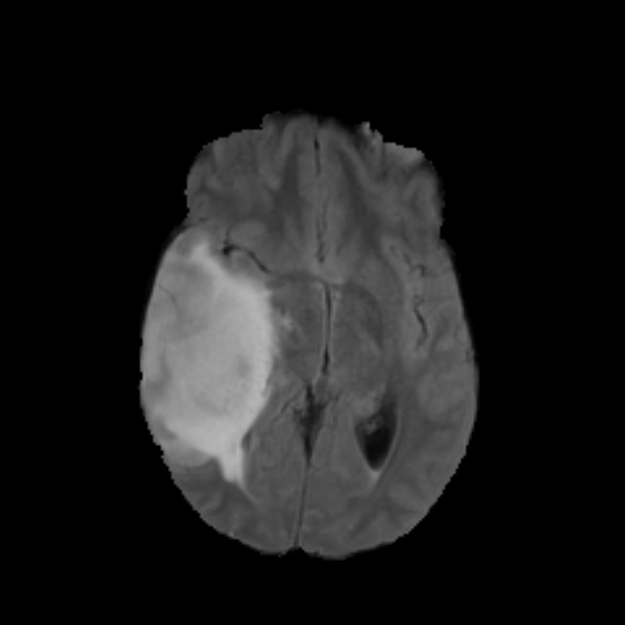} &
\includegraphics[width=0.13\textwidth]{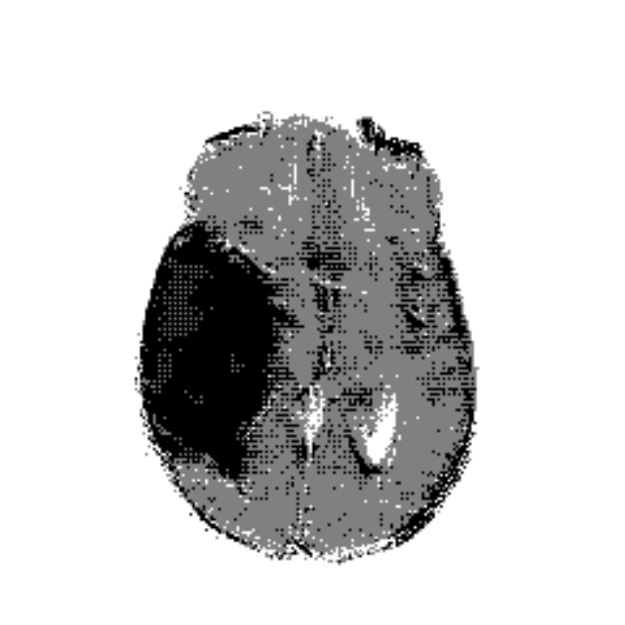} &
\includegraphics[width=0.13\textwidth]{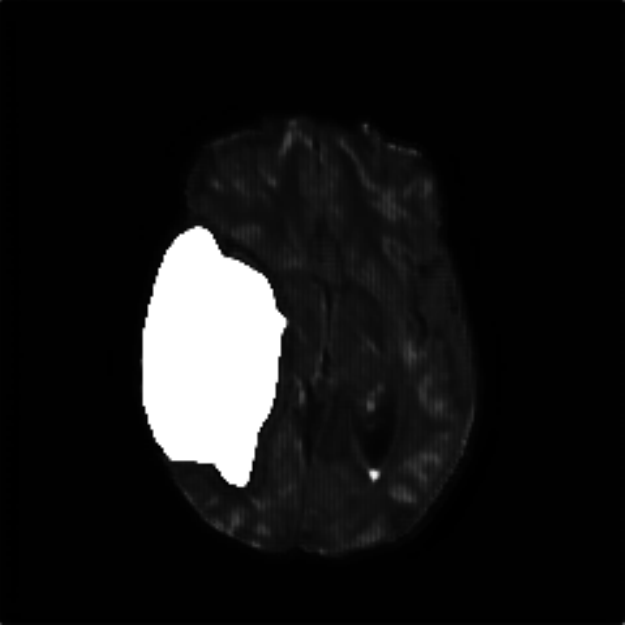} &
\includegraphics[width=0.13\textwidth]{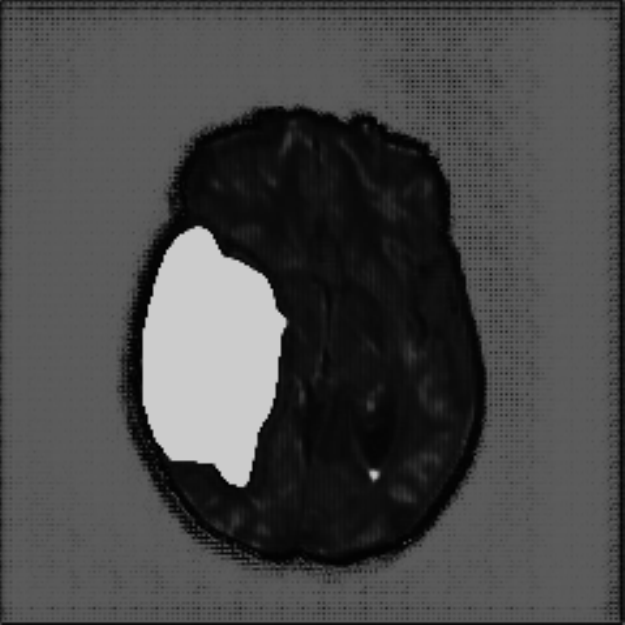} &
\includegraphics[width=0.13\textwidth]{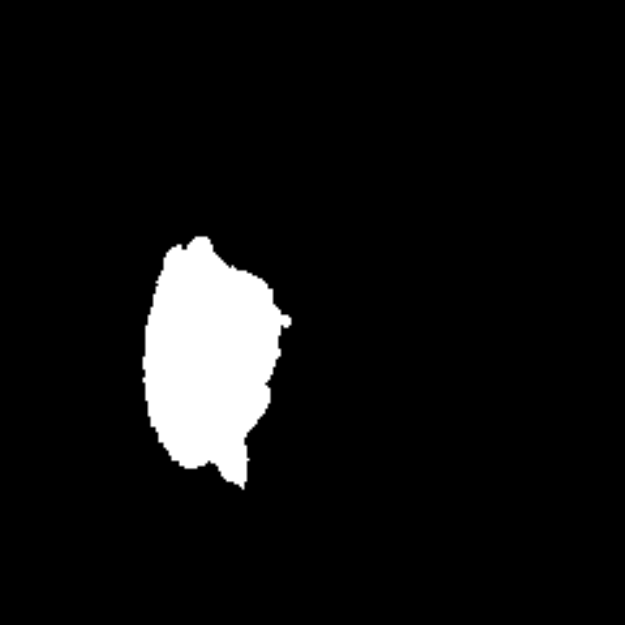} &
\includegraphics[width=0.13\textwidth]{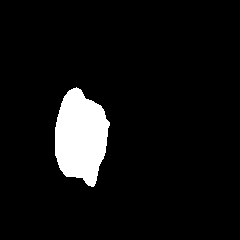} &
\includegraphics[width=0.13\textwidth]{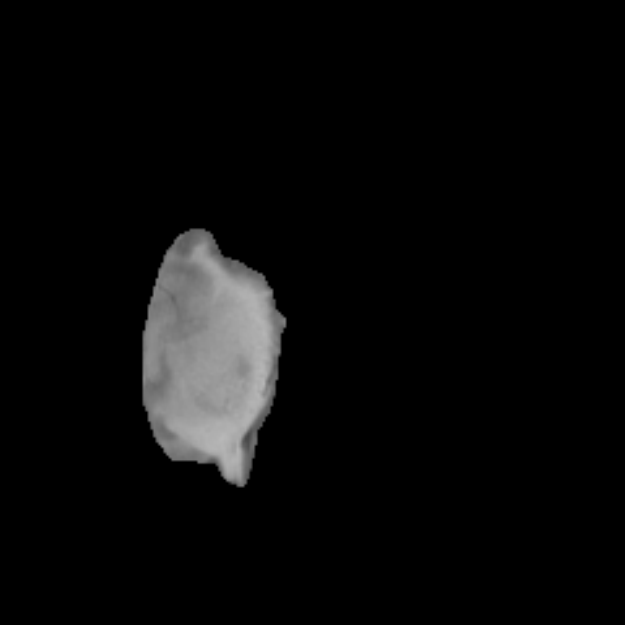} \\

\includegraphics[width=0.13\textwidth]{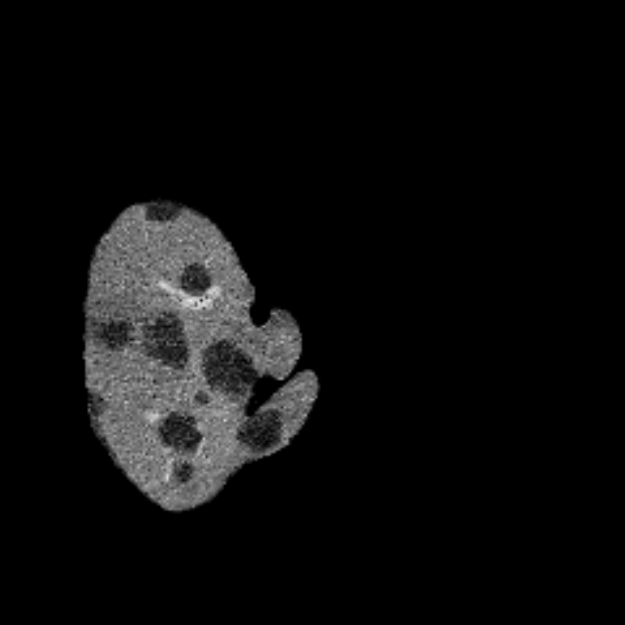} &
\includegraphics[width=0.13\textwidth]{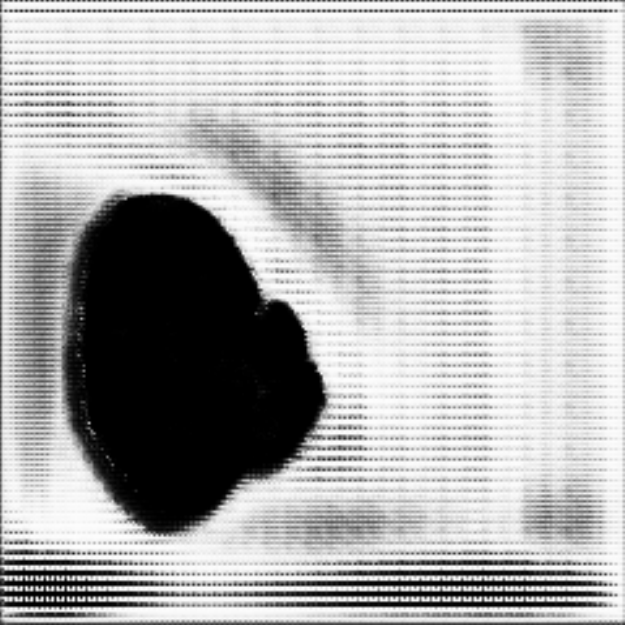} &
\includegraphics[width=0.13\textwidth]{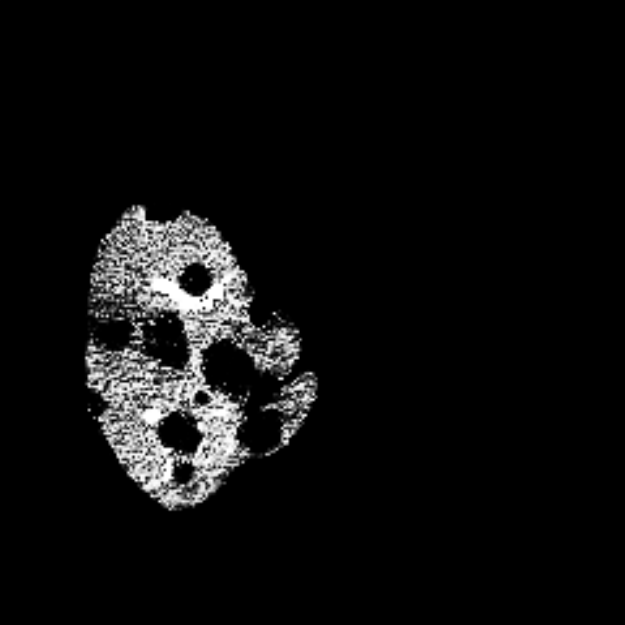} &
\includegraphics[width=0.13\textwidth]{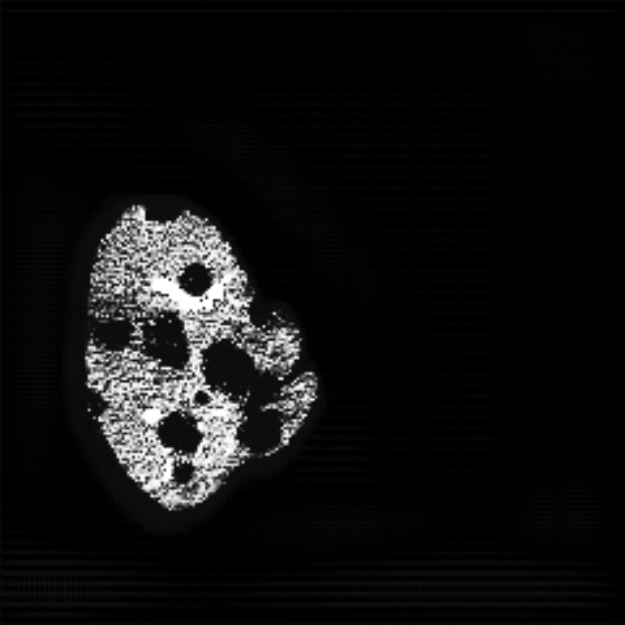} &
\includegraphics[width=0.13\textwidth]{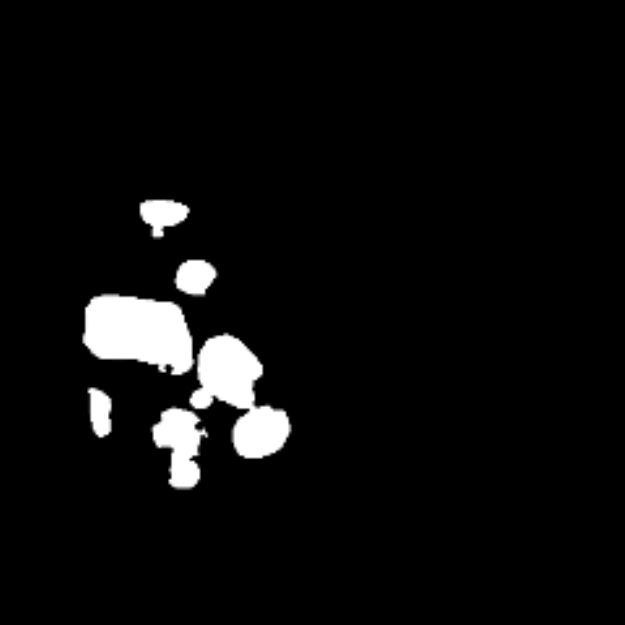} &
\includegraphics[width=0.13\textwidth]{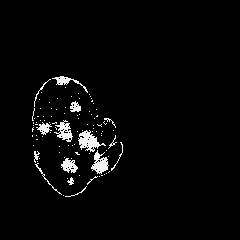} &
\includegraphics[width=0.13\textwidth]{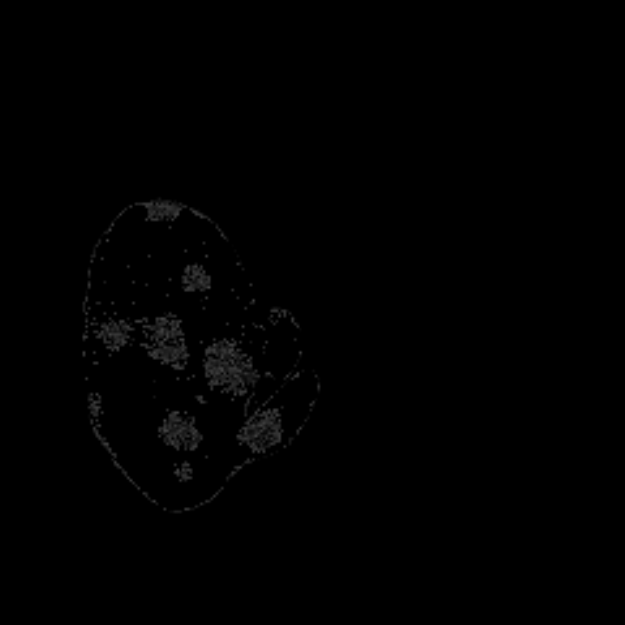} \\
\includegraphics[width=0.13\textwidth]{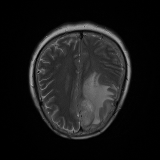} &
\includegraphics[width=0.13\textwidth]{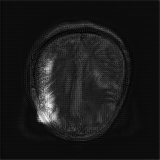} &
\includegraphics[width=0.13\textwidth]{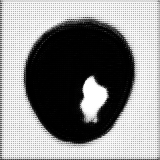} &
\includegraphics[width=0.13\textwidth]{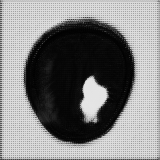} &
\includegraphics[width=0.13\textwidth]{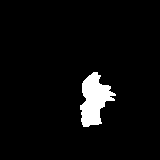} &
\includegraphics[width=0.13\textwidth]{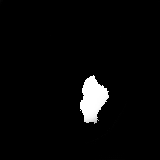} &
\includegraphics[width=0.13\textwidth]{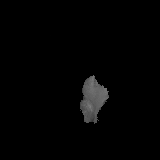}
\end{tabular}
\caption{Sample results of MS-SEG2015, Brats-2019, LiTS, and our private dataset (top to bottom) obtained from the various branches of our ASC-Net. Left to right: the input image $I_{in}$, the output image of the fence cut $I_{fc}$, which is contrast enhanced to present the content contained in the brain region, the output image of the wild cut $I_{wc}$, the reconstructed image $I_{ro}$, the ground-truth mask $M_{gt}$, our predicted segmentation mask $M_{pred}$, and the predicted region of interst $M_{pred}\cap I_{in}$. None of these include any of the post-processed images.} 
\label{fig:result_brats}
\end{figure*}

\section{Applications}

We evaluate our model on three unsupervised anomaly segmentation tasks: MS lesion segmentation, brain tumor segmentation, and liver lesion segmentation. We use the MS-SEG2015~\cite{carass2017longitudinal} training set,  BraTS~\cite{bakas2017advancing,bakas2018identifying,menze2014multimodal}, and  LiTS~\cite{bilic2019liver} datasets in these tasks. Also, a private dataset is collected to further evaluate the effectiveness of our method.  


\subsection{Datasets and Experimental Settings}

\vspace{0.05in}
\noindent
\textbf{MS-SEG2015.} The training set consists of 21 scans from 5 subjects with each scan dimensions of $181 \times217 \times 181$. We resize the axial slices to $160 \times 160$, so that we can share the same network design as the rest of the experiments. 

\vspace{0.05in}
\noindent
\textbf{BraTS 2019.} This dataset consists of 335 MRI brain scans collected from 259 subjects with High Grade Glioma (HGG) and 76 subjects with Low Grade Gliomas (LGG) in the training set. The 3D dimensions of the images are $240 \times 240 \times 155$. 

\vspace{0.05in}
\noindent
\textbf{LiTS.} The training set of LiTS consists of 130 abdomen
CT scans of patients with liver lesions, collected from multiple institutions. Each scan has a varying number of slices with dimensions of 512$\times$512. We resize these CT slices to $240 \times 240$ to share the same network architecture with other tasks. 

\vspace{0.05in}
\noindent
\textbf{Private Dataset.} The in-house dataset collected from Zhongnan Hospital in Wuhan, China consists of non-skull-stripped brain MRI images of different modalities, including T2 FSE (Fast Spin Echo, short for T2) and T2 Flair (short for Flair) scans. In this dataset, we have 55 normal control subjects for establishing the reference distribution, 26 subjects with brain tumors for training, and 41 subjects with brain tumors and manually-segmented masks for testing. 
The images are all of different resolutions and thus we resample them to $160\times160$ resolution on 2D slices. 


\vspace{0.05in}
\noindent
For all experiments, the image intensity is normalized to $[0, 1]$ over the 3D volume; however, we perform the 3D segmentation task in the slice-by-slice manner using axial slices. To balance the sample size in $I_{in}$ and $R_d$, we randomly sample and duplicate the number difference to the respective set.

\begin{table}[t]
\centering
\begin{tabular}{l|cc}
 Model & w/o post (\%) & w post (\%) \\ 
 \hline
AE(dense)  & 4.5 & 15.0$\pm$7.5   \\
Context AE  & 5.1 & 18.8$\pm$11.6  \\
VAE  & 5.1 & 20.0$\pm$12.4   \\
Context VAE (original)  & 6.0 & 26.7$\pm$11.2  \\
Context VAE (gradient)  & 5.9 & 12.7$\pm$8.8   \\
GMVAE (dense)  & 6.3 & 17.4$\pm$12.1   \\
GMVAE (dense restoration)  & 6.3 & 22.3$\pm$12.4   \\
GMVAE (spatial)  & 2.8 & 6.9$\pm$7.3   \\
GMVAE (spatial restoration)  & 2.7 & 11.8$\pm$11.0   \\
f-AnoGAN  & 8.9 & \textbf{\color{blue}{27.8$\pm$14.0}}   \\
Constrained AE  & \textbf{\color{blue}{9.7}} & 20.9 $\pm$10.0   \\
Constrained AAE  & 6.8 & 19.0$\pm$17.0   \\
VAE (restoration)  & 5.8 & 21.1$\pm$12.2   \\
AnoVAEGAN  & 6.6 & 20.0$\pm$13.3  \\
Ours             & \textbf{32.94$\pm$35.98} & \textbf{48.20$\pm$47.84}  \\
\end{tabular}
\caption{Experimental comparison of anomaly segmentation on MS-SEG2015 across different methods via experiments conducted in~\cite{baur2021autoencoders} and ours. The best results reported in~\cite{baur2021autoencoders} are colored in \textbf{\color{blue}{blue}}.}
\label{tab:MS-Seg-res}
\end{table}

\begin{table*}[t]
\centering
\begin{tabular}{c|cc|ccc}
\multirow{2}{*}{Dice (\%)}& \multicolumn{2}{c|}{Public Datasets} &  \multicolumn{3}{c}{Private Dataset$^*$} \\

 &BraTS 2019 & LiTS  & Flair (CV)  & T2 (CV)& T2 (Full)\\ 
 \hline
w/o post-processing  & 63.67$\pm$16.24 & 32.24$\pm$20.74 & 79.89$\pm$49.32  & 88.57$\pm$46.28  & 85.79$\pm$47.63\\
w post-processing  & 68.01$\pm$14.63 & \textbf{50.23$\pm$32.22} & -- & \textbf{91.58$\pm$46.81} & \textbf{90.21$\pm$46.37}  \\
\hline
best baseline & \textbf{71.63$\pm$0.84$^{\dagger}$}~\cite{zhang2021self} & 40.78$\pm$0.43~\cite{zhang2021self}& -- & --  &  -- 
\end{tabular}
\caption{Experimental comparison of anomaly segmentation on remaining two public datasets and one private dataset. $^*$For post processing the segmentation outputs of the private dataset, we do not use the open-and-closed operation, because the predicted segmentation masks already consist of connected regions and such post-processing would make the result worse. Instead, we use the Flair modality to remove the false positives, like the CSF regions, which is easily mixed with tumors in the T2 modality based on our observations.$^\dagger$Their result was reported on BraTS 2018. See detaled discussion in Section~\ref{sec:exp_res}. CV is short for cross validation, and ``Full" indicates using the entire private dataset.}
\label{tab:three-datasets-results}
\end{table*}

\subsection{Experimental Results}
\label{sec:exp_res}
\noindent
\textbf{MS Lesion Segmentation (MS-SEG2015).} In this task, we randomly sample $2870$ non-tumor, non-zero, Brats-2019 training set slices to make our reference distribution $R_d$ as in~\cite{baur2021autoencoders}, while they use their own privately annotated healthy dataset. Meanwhile, the $2870$ non zero 2D slices of the MS-SEG2015 training set are used in the main module $M$. We train this network using three cycles in the first stage and one cycle in the second stage and take the threshold at 254 intensity\footnote{The intensity range for computing the image histogram is [0, 255].} based on the right most peak of the image histogram. 

We obtain a subject-wise mean Dice score of 32.94\% without any post-processing. By using a simple post-processing with erosion and dilation with $5\times5$ filters, this number improves to 48.20\% mean Dice score. In comparison, a similar study conducted by~\cite{baur2021autoencoders} consisting of a multitude of algorithms including AnoVAEGAN \cite{baur2018deep} and f-AnoGANS, obtained a best mean score of 27.8\% Dice after post-processing by f-AnoGANS. Before post-processing the best method was Constrained AutoEncoder~\cite{chen2018unsupervised} with a score of 9.7\% Dice. An exhaustive list is presented in Table~\ref{tab:MS-Seg-res}. Figure~\ref{fig:result_brats} shows sample images of our results.


\vspace{0.05in}
\noindent
\textbf{Brain Tumor Segmentation (BraTS 2019).} In this task, we perform patient-wise two-fold cross-validation on the Brats-2019 training set. In each training fold, we use a 90/10 split after removing empty slices. The 2D slices from the 90\% split without tumors are used to make our reference distribution $R_d$; while the 2D slices with tumors from the 90\% split and all the slices from the 10\% split are used for training our model. As a result, the sample size of $R_d$ for fold one and two amounts to 11,745 and 12,407 respectively, while the size of $I_{in}$ amounts to 11,364 and 10,786, respectively. We train this network using two cycles in the first stage and one cycle in the second stage.






\begin{figure*}[t]
\centering
\includegraphics[width=0.96\textwidth]{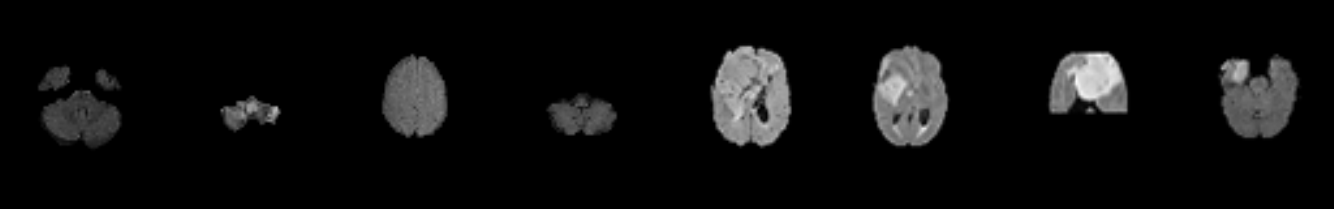}
\includegraphics[width=0.96\textwidth]{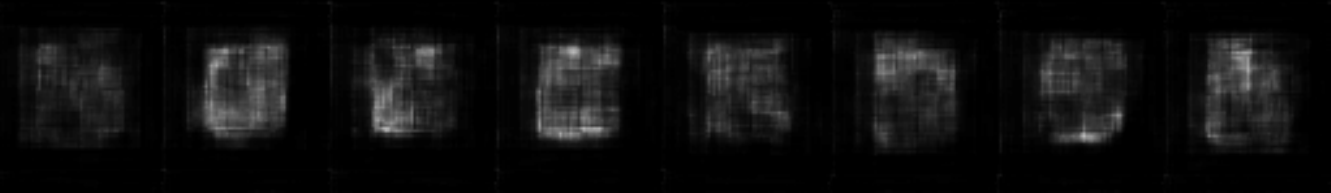}
\caption{Query images (top) and their reconstructions (bottom) using f-AnoGANs~\cite{schlegl2019f}. }
\label{fig:anogans_fail}
\end{figure*}

\begin{figure*}[t]
\centering
\setlength\tabcolsep{1.5pt}
\begin{tabular}{cccccccc}
$Flair_{in}$ & $Flair_{ro}$ & ${Flair_{pred}}$ &$T2_{in}$& $T2_{ro}$& ${T2_{pred}}$ & $M_{T2*Fl}$ & $M_{gt}$ \\
\includegraphics[width=0.117\textwidth]{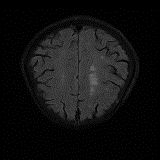} &
\includegraphics[width=0.117\textwidth]{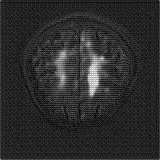} &
\includegraphics[width=0.117\textwidth]{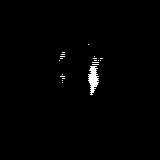} & 
\includegraphics[width=0.117\textwidth]{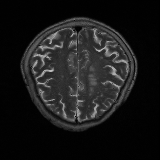}&
\includegraphics[width=0.117\textwidth]{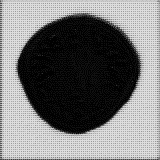} &
\includegraphics[width=0.117\textwidth]{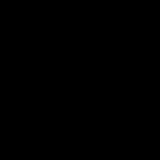}& 
\includegraphics[width=0.117\textwidth]{Final_image_set/T2_vs_Flair/t0_t.png}& 
\includegraphics[width=0.117\textwidth]{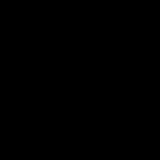} 
 \\
 \includegraphics[width=0.117\textwidth]{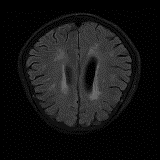}  &
\includegraphics[width=0.117\textwidth]{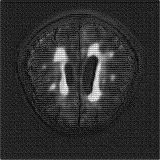} &
\includegraphics[width=0.117\textwidth]{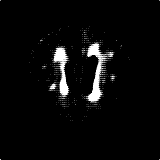} &
\includegraphics[width=0.117\textwidth]{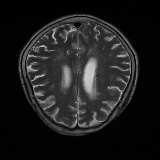}&
\includegraphics[width=0.117\textwidth]{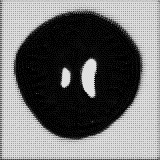} &
\includegraphics[width=0.117\textwidth]{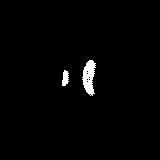}&
\includegraphics[width=0.117\textwidth]{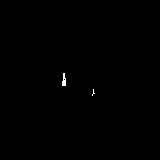} &
\includegraphics[width=0.117\textwidth]{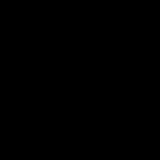}
 \\
 \includegraphics[width=0.117\textwidth]{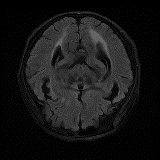}  &
\includegraphics[width=0.117\textwidth]{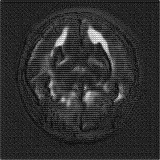} &
\includegraphics[width=0.117\textwidth]{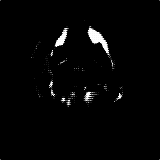} &
\includegraphics[width=0.117\textwidth]{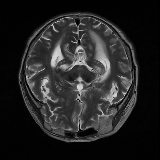}&
\includegraphics[width=0.117\textwidth]{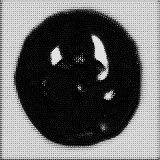} &
\includegraphics[width=0.117\textwidth]{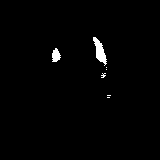}&
\includegraphics[width=0.117\textwidth]{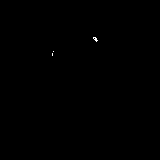} &
\includegraphics[width=0.117\textwidth]{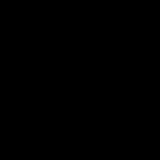}
\\
 \includegraphics[width=0.117\textwidth]{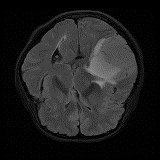} &
\includegraphics[width=0.117\textwidth]{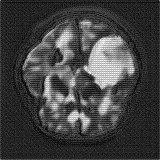} &
\includegraphics[width=0.117\textwidth]{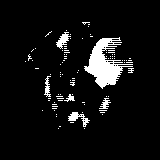} &
\includegraphics[width=0.117\textwidth]{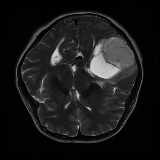}&
\includegraphics[width=0.117\textwidth]{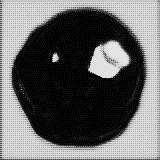} &
\includegraphics[width=0.117\textwidth]{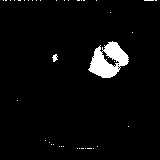}&
\includegraphics[width=0.117\textwidth]{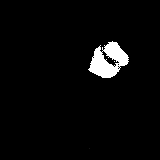} &
\includegraphics[width=0.117\textwidth]{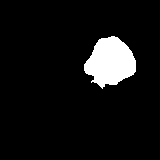} 
\end{tabular}
\caption{Result comparison between T2 Flair (short for Flair) and T2 FSE (short for T2), Part I. Using the Flair scans has the potential of detecting additional anomalies; while using the T2 scans would mix tumors with other regions like CSF (the last three rows), which can be cleaned by simply using the flair input image to post process the T2 predictions (the second to the last column). Left to right: the Flair input image ($Flair_{in}$ ), the reconstructed Flair input ($Flair_{ro}$), the predicted segmentation mask on the Flair input (${Flair_{pred}}$), the T2 input image ($T2_{in}$), the reconstructed T2 input ($T2_{ro}$), the predicted segmentation mask on the T2 input (${T2_{pred}}$), the predicted segmentation mask after post-processing by using the Flair input ($M_{T2*Fl}$), and the ground-truth mask ($M_{gt}$).}
\label{fig:Flair_vs_t2_1}
\end{figure*}

\begin{figure*}[t]
\centering
\setlength\tabcolsep{1.5pt}
\begin{tabular}{cccccccc}
$Flair_{in}$ & $Flair_{ro}$ & ${Flair_{pred}}$ &$T2_{in}$& $T2_{ro}$& ${T2_{pred}}$ & \small{$M_{T2*Fl}$} & $M_{gt}$\\
 \includegraphics[width=0.117\textwidth]{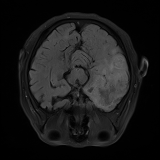}  &
\includegraphics[width=0.117\textwidth]{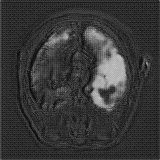} &
\includegraphics[width=0.117\textwidth]{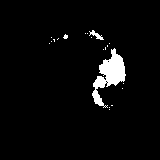} &
\includegraphics[width=0.117\textwidth]{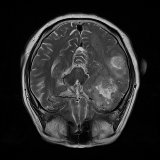}&
\includegraphics[width=0.117\textwidth]{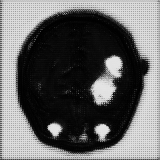} &
\includegraphics[width=0.117\textwidth]{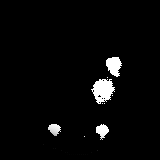}&
\includegraphics[width=0.117\textwidth]{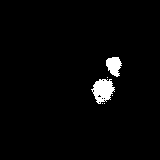} &
\includegraphics[width=0.117\textwidth]{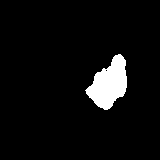}
\\
 \includegraphics[width=0.117\textwidth]{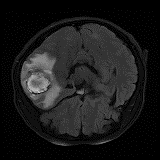} &
\includegraphics[width=0.117\textwidth]{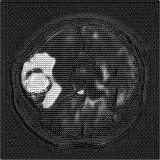} &
\includegraphics[width=0.117\textwidth]{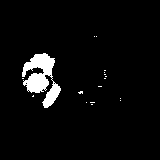} &
\includegraphics[width=0.117\textwidth]{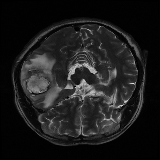}&
\includegraphics[width=0.117\textwidth]{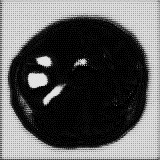} &
\includegraphics[width=0.117\textwidth]{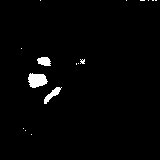} &
\includegraphics[width=0.117\textwidth]{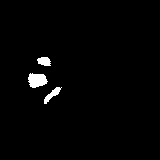} &
\includegraphics[width=0.117\textwidth]{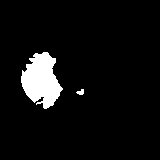} 
 \\
 \includegraphics[width=0.117\textwidth]{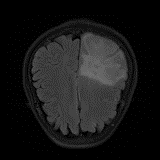} &
\includegraphics[width=0.117\textwidth]{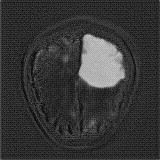} &
\includegraphics[width=0.117\textwidth]{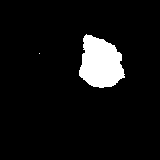} &
\includegraphics[width=0.117\textwidth]{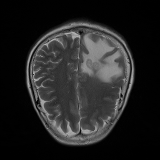}&
\includegraphics[width=0.117\textwidth]{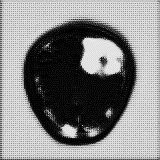} &
\includegraphics[width=0.117\textwidth]{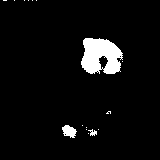}&
\includegraphics[width=0.117\textwidth]{Final_image_set/T2_vs_Flair/t6_t.png} &
\includegraphics[width=0.117\textwidth]{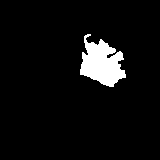} 
 \\
 \includegraphics[width=0.117\textwidth]{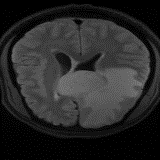} &
\includegraphics[width=0.117\textwidth]{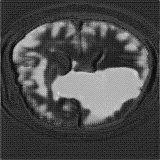} &
\includegraphics[width=0.117\textwidth]{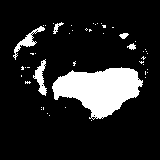} &
\includegraphics[width=0.117\textwidth]{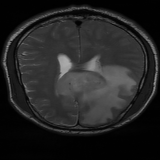}&
\includegraphics[width=0.117\textwidth]{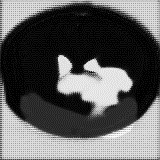} &
\includegraphics[width=0.117\textwidth]{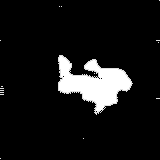} &
\includegraphics[width=0.117\textwidth]{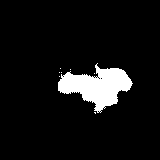} &
\includegraphics[width=0.117\textwidth]{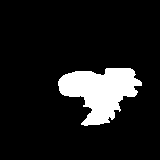} 
\end{tabular}
\caption{Result comparison between T2 Flair (short for Flair) and T2 FSE (short for T2), Part II. Using the Flair scans suffers the issues of under segmentation (the first two rows) and over segmentation (the last two rows); while using the T2 scans often suffers the under-segmentation issue, since the over-segmentation issue can be mostly handled by post-processing similar to Fig.~\ref{fig:Flair_vs_t2_1}. Besides, using the Flair image usually obtains a better tumor segmentation with cleaner edges compared to the T2 sample. Left to right: the Flair input image ($Flair_{in}$ ), the reconstructed Flair input ($Flair_{ro}$), the predicted segmentation mask on the Flair input (${Flair_{pred}}$), the T2 input image ($T2_{in}$), the reconstructed T2 input ($T2_{ro}$), the predicted segmentation mask on the T2 input (${T2_{pred}}$), the predicted segmentation mask after post-processing by using the Flair input ($M_{T2*Fl}$), and the ground-truth mask ($M_{gt}$).}
\label{fig:Flair_vs_t2_2}
\end{figure*}

\begin{figure*}[t]
\centering
\setlength\tabcolsep{1.5pt}
\begin{tabular}{cccccccc}
$Flair_{in}$ & $Flair_{ro}$ & ${Flair_{pred}}$ &$T2_{in}$& $T2_{ro}$& ${T2_{pred}}$ & \small{$M_{T2*Fl}$} & $M_{gt}$\\
\includegraphics[width=0.117\textwidth]{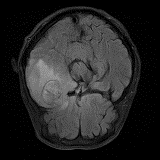} &
\includegraphics[width=0.117\textwidth]{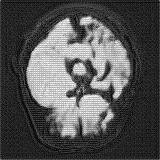} &
\includegraphics[width=0.117\textwidth]{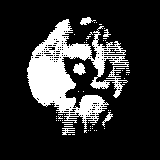} &
\includegraphics[width=0.117\textwidth]{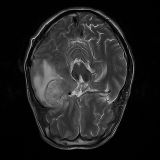}&
\includegraphics[width=0.117\textwidth]{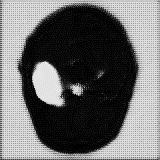} &
\includegraphics[width=0.117\textwidth]{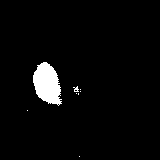}&
\includegraphics[width=0.117\textwidth]{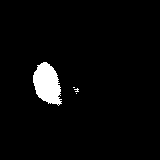}  &
\includegraphics[width=0.117\textwidth]{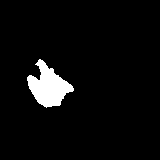}
 \\
 \includegraphics[width=0.117\textwidth]{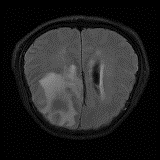} &
\includegraphics[width=0.117\textwidth]{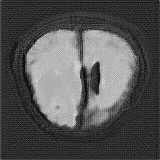} &
\includegraphics[width=0.117\textwidth]{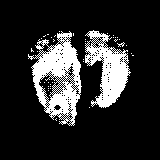} &
\includegraphics[width=0.117\textwidth]{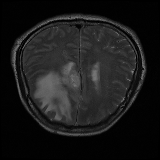}&
\includegraphics[width=0.117\textwidth]{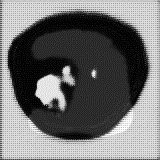} &
\includegraphics[width=0.117\textwidth]{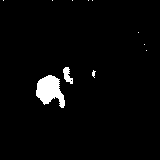}&
\includegraphics[width=0.117\textwidth]{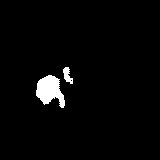}&
\includegraphics[width=0.117\textwidth]{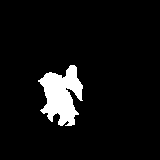}  
\\
 \includegraphics[width=0.117\textwidth]{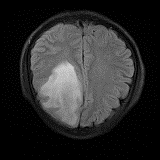} &
\includegraphics[width=0.117\textwidth]{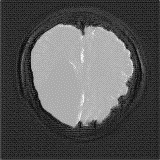} &
\includegraphics[width=0.117\textwidth]{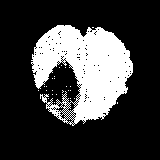} &
\includegraphics[width=0.117\textwidth]{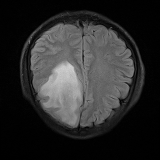}&
\includegraphics[width=0.117\textwidth]{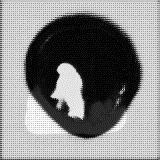} &
\includegraphics[width=0.117\textwidth]{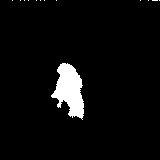}&
\includegraphics[width=0.117\textwidth]{Final_image_set/T2_vs_Flair/t7_t.png}&
\includegraphics[width=0.117\textwidth]{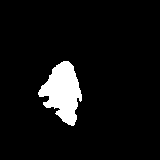} 
\end{tabular}
\caption{Result comparison between T2 Flair (short for Flair) and T2 FSE (short for T2), Part III. Using the Flair scans fails to detect some anomalies; while using the T2 scans can consistently obtain the rough locations of the anomalies, although it still suffers the under segmentation issue. We also see a cluster flip for the Flair case in row 3, where the tumor is part of the left most peak and analysis of these kinds of flips is left for future works. Left to right: the Flair input image ($Flair_{in}$ ), the reconstructed Flair input ($Flair_{ro}$), the predicted segmentation mask on the Flair input (${Flair_{pred}}$), the T2 input image ($T2_{in}$), the reconstructed T2 input ($T2_{ro}$), the predicted segmentation mask on the T2 input (${T2_{pred}}$), the predicted segmentation mask after post-processing by using the Flair input ($M_{T2*Fl}$), and the ground-truth mask ($M_{gt}$). }
\label{fig:Flair_vs_t2_3}
\end{figure*}

We obtain a subject-wise mean Dice score of 63.67\% for the brain tumor segmentation. Utilizing a simple post-processing scheme of erosion and dilation with $9 \times 9$ filter, we improve our mean Dice score to 68.01\%. 
Figure~\ref{fig:result_brats} shows samples generated by our ASC-Net and Table~\ref{tab:three-datasets-results} shows our before and after post-processing results. We attempted to apply f-AnoGANs~\cite{schlegl2019f} by following their online instructions and failed to generate good reconstructions as shown in Figure~\ref{fig:anogans_fail}. The failure of AnoGANs in the reconstruction brings to light the issue with the regeneration based methods and the complexity and stability of GAN-based image reconstruction. 

A most recent work in~\cite{naval2021implicit} trains their algorithm using 1,112 healthy scans from the Human Connectome Project (HCP) young adult dataset~\cite{van2012human} and tests on 50 random BraTS 2018 scans, obtaining a mean dice score of 67.2\% and 15.5\% standard deviation. Following our simple post-processing scheme, our algorithm performs better, increasing the mean Dice score by 0.81\% and reducing the standard deviation by 0.97\%, on two-fold cross validation across 335 scans. Another work in~\cite{zhang2021self} tests on the BraTS 2018 training set, obtaining a mean dice score of 71.63\% and standard deviation of 0.84\%. While their method outperforms ours, it is worth to mention that the self-supervised method is highly specialized to a particular task of tumor segmentation. It may happen that the object to be segmented is difficult to synthesize artificially or perfectly, resulting in a bottleneck of the pipeline. Furthermore, one assumption of a self-supervised learning algorithm is that the object to be learnt is known beforehand. Thus, a model trained for brain tumors cannot be applied readily to other anomalies, e.g., brain lesions. Our method, on the other hand, has no such limitations and does not need Pseudo dataset generation for a new task. That is, our method is a general approach for anomaly detection and segmentation. Also, our method performs better than~\cite{zhang2021self} on the liver lesion segmentation task after post-processing.



\vspace{0.05in}
\noindent
\textbf{Liver Lesion Segmentation (LiTS).} To generate the image data for this task, we remove the non-liver region by using the liver mask generated by CompNet~\cite{dey2018compnet} and take all non-zero images. We have 11,926 2D slices without liver lesions used in the reference distribution $R_d$. The remaining 6,991 images is then used for training the model. We perform slice-by-slice two-fold cross-validation and train the network using two cycles in both first and second stages. To extract the liver lesions, we first mask out the noises in the non-liver region of the reconstructed image $I_{ro}$ and then invert the image to take a threshold value at 242, the rightmost peak of the inverted image.

We obtain a slice-wise mean Dice score of 32.24\% for this liver lesion segmentation, 
which improves to 50.23\% by using a simple post processing scheme of erosion and dilation with $5\times5$ filter.  Sampled results are shown in Fig.~\ref{fig:result_brats}. Compared with~\cite{zhang2021self}, which obtains a mean Dice score of 40.78\% and a standard deviation of 0.43\%, we improve the mean Dice score by almost 10\%, but has a much larger standard deviation. Unlike~\cite{zhang2021self}, where the network is pre-trained on a artificial tumor dataset, and hence the pipeline customized for tumor segmentation, our method do not need such information beforehand. We notice that our standard deviation for BraTS dataset is similar to~\cite{naval2021implicit}. This is because novelty/anomaly detection algorithms without a pre-defined task would suffer from the co-morbidities issues discussed in Section~\ref{limitations}.

Our method still has room to improve, compared with supervised methods. A recent study~\cite{dey2020hybrid} reports a cross-validation result of 67.3\% under a supervised setting. 
Note that the annotation in the LiTS lesion dataset is imperfect with missing small lesions~\cite{chlebus2018automatic,dey2020hybrid}. Since we use the imperfect annotation to select images for the reference distribution, some slices with small lesions may be included and treated as normal examples. 
Also, the faint liver boundary causes a fixed penalty to be incurred per slice, which reduces the dice score. This kind of segmentation noise could be better handled by using a more sophisticated post-processing strategy. 


\vspace{0.05in}
\noindent
\textbf{Brain Tumor Segmentation (Private Dataset without Skull-Stripping).} In this dataset we have access to good quality reference distributions on different modalities; therefore, we further investigate the performance of our method using T2 and Flair modalities. We perform two-fold cross-validation experiments on those 41 subjects with annotations, using firstly the Flair modality and then followed by T2. In this setting, we do not use the images from the 26 subjects without tumor masks. The reference distribution for each experiment consists of image slices collected from the 55 normal control patients.

The Flair experiment results are obtained using a training of two cycles in the first stage and four cycles in the second stage to reach the peak separation, and the threshold value is taken at the intensity of 170. For T2 we use two cycles in the first stage and two cycles in the second stage, and the threshold is taken at 220. Both thresholds are taken based on the rightmost peaks of the histograms of reconstructed images. We obtain a subject-wise mean dice score of 79.89\% on the Flair scans and 88.57\% on the T2 scans. Despite the lower score compared to T2, the Flair modality provides the potential of identifying additional anomalies, which may not be limited to HGG or LGG, as shown in the third column of Fig.~\ref{fig:Flair_vs_t2_1}. 
However, since the focus of this experiment is to segment HGG and LGG only, using the T2 modality outperforms the Flair in term of the Dice scores as reported in Table~\ref{tab:three-datasets-results}, and the predicted masks as shown in Fig.~\ref{fig:Flair_vs_t2_1}. Aside from that, on the Flair scans our method suffers both under- and over-segmentation as shown in Fig.~\ref{fig:Flair_vs_t2_2} and struggles to segment tumors using one uniform threshold as shown in Fig.~\ref{fig:Flair_vs_t2_3}. Typically, we use the rightmost peak as the threshold for brighter tumors; however, the peaks separating tumors in these cases of Fig.~\ref{fig:Flair_vs_t2_3} occur as the leftmost peak. Such flip further lowers the segmentation score, even though the algorithm is able to separate the anomaly as one of the two cuts. 

Although our method has consistently better performance of segmenting brain tumors on T2, as shown from Fig.~\ref{fig:Flair_vs_t2_1} to Fig.~\ref{fig:Flair_vs_t2_3}.
In the case of T2, the primary disadvantages occur due to the inclusion of other regions, such as Cerebrospinal Fluid (CSF), eyeballs, etc., which appear dark in the Flair modality. In order to alleviate these false positives on T2 scans, we multiply the predicted masks with the Flair input images. Then we re-calibrate the output by taking a threshold at the intensity of 50 (roughly 0.2 in the range [0,1]) to generate our final mask. This post-processing is our new choice for the private dataset. We did not use the erosion/dilation operation because it is more efficient for cases with discontinuous segmentation results, which is the one our public datasets suffer with, as shown in Fig.~\ref{fig:result_brats}, but not our private dataset. This new post-processing improves the performance to a patient-wise mean dice score of 91.58\% on T2 scans with two-fold cross validation.




Since the two-fold cross-validation experiment highlights the strengths using the T2 modality with respect to segmenting HGG and LGG, we next perform a third experiment, utilizing the entire dataset with T2 scans. In this task, the reference distribution is composed of 1224 2D slices obtained from 55 patients, and the training input image passed to the main module is composed of 846 2D slices from 26 subjects without annotations. The training input images consist of a mixture of patients with and without tumors. The test set similarly consists of patients with and without tumors from 41 annotated subjects with 1151 2D slices. We train our model using two cycles in the first stage and one cycle in the second stage. The threshold value of 220 is selected based on the rightmost peak of the image histogram, resulting in a subject-wise mean dice score of 85.79\%. As reported in Table~\ref{tab:three-datasets-results}, we achieve a much better result compared to the similar task on BraTS, since we have a good reference distribution curated from scans without anomaly present and also good segmentation masks for evaluation. our post-processing step for the private dataset further improves the performance to a patient-wise mean dice score of 90.21\%.

\begin{figure*}[h]
\centering
  \includegraphics[width=0.115\textwidth]{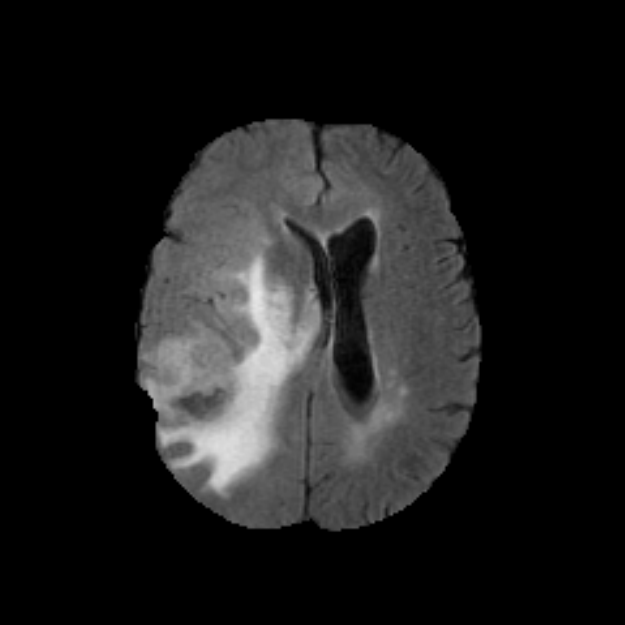}
  \includegraphics[width=0.115\textwidth]{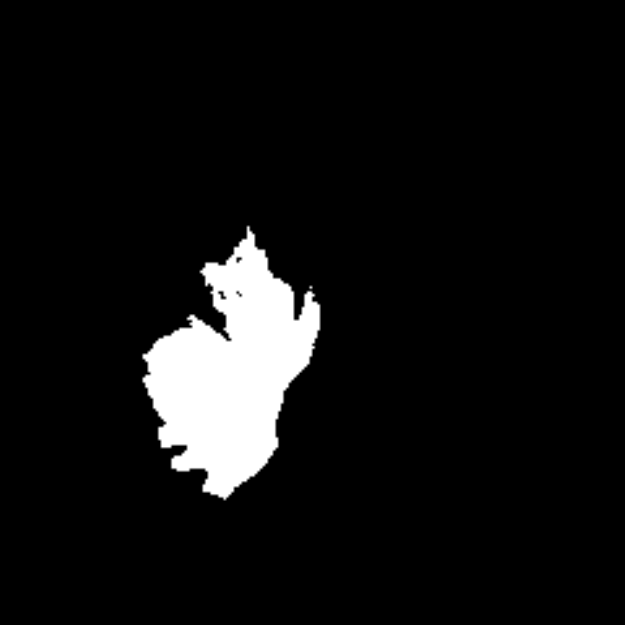}
  \includegraphics[width=0.115\textwidth]{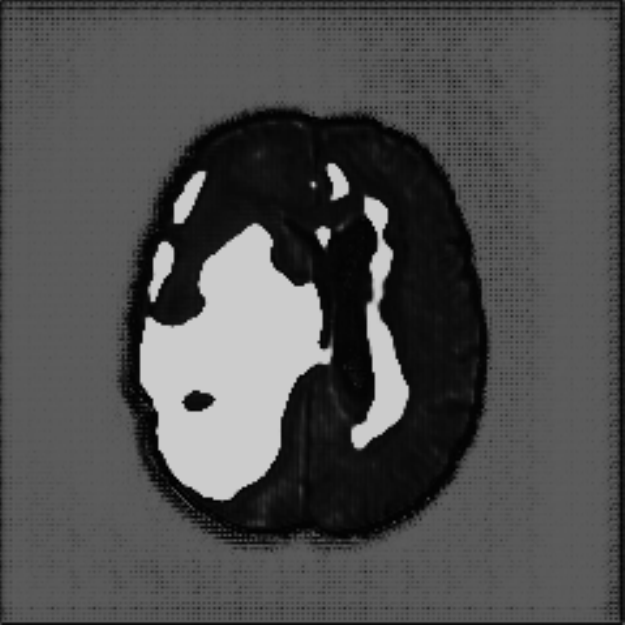}
    \includegraphics[width=0.115\textwidth]{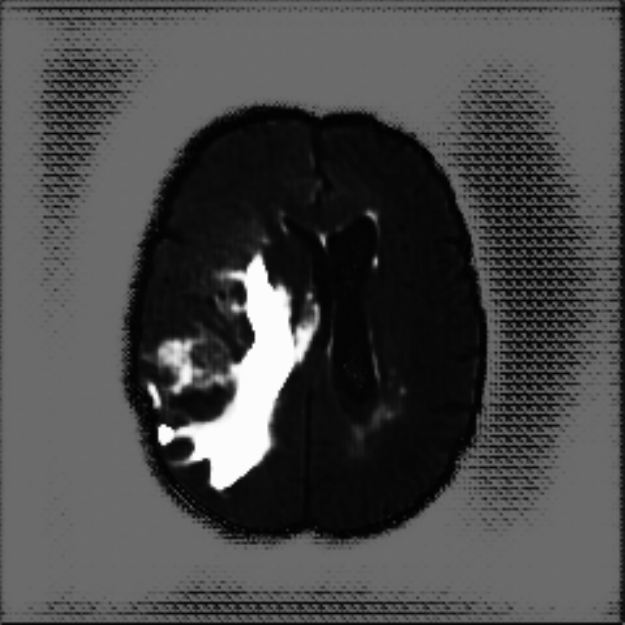}
  \includegraphics[width=0.115\textwidth]{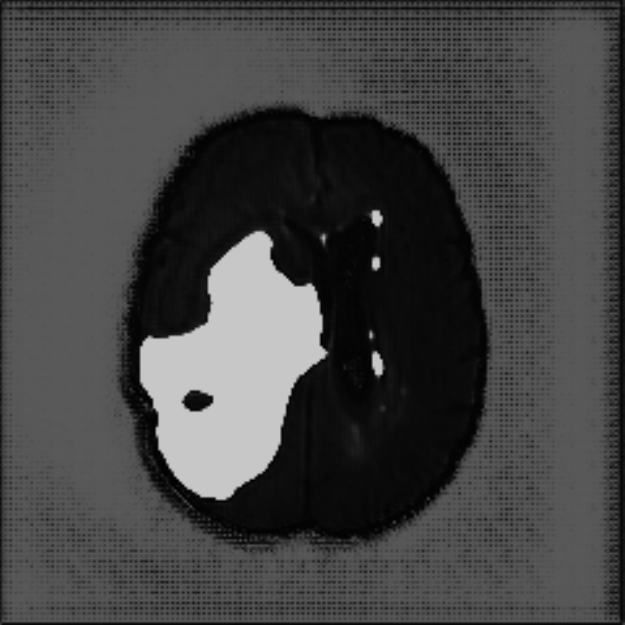}
    \includegraphics[width=0.115\textwidth]{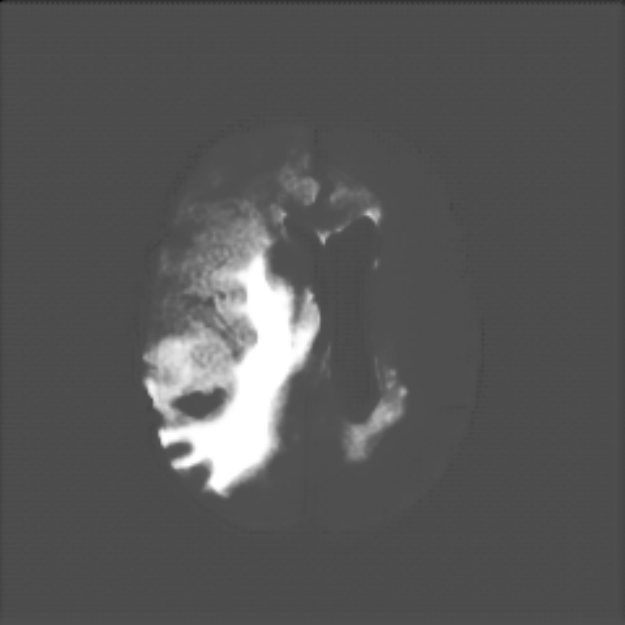}
  \includegraphics[width=0.115\textwidth]{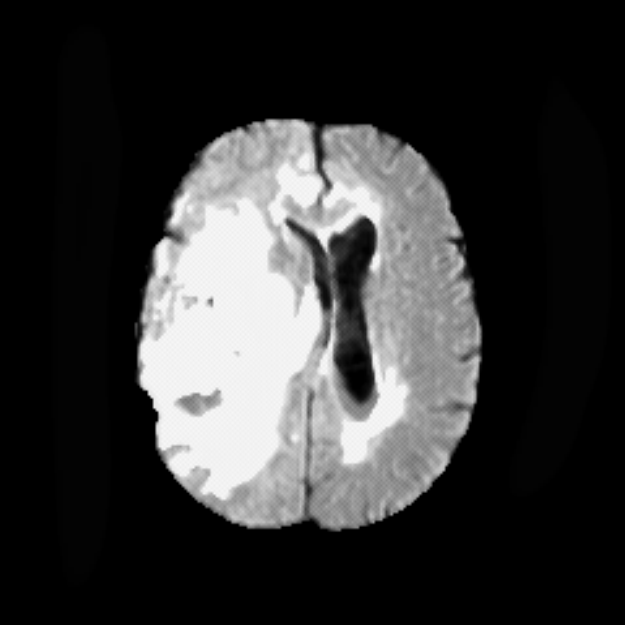}
  \includegraphics[width=0.115\textwidth]{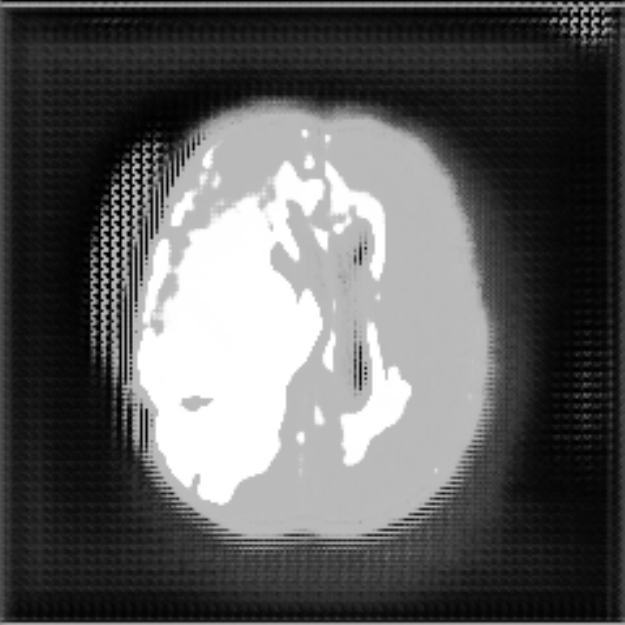}
\caption{Stability: The first image is the input image, the second is the ground truth. The rest of images are reconstruction from various re-runs of the framework with variable training cycles and stage. All runs are able to isolate the anomaly in question.}
\label{fig:stability}
\end{figure*}

\section{Discussion and Future Work}
In this paper we have presented a framework that performs two-cut split in an unsupervised fashion guided by an reference distribution using GANs. Unlike the methods in AnoGAN's family, our ASC-Net focuses on the anomaly detection with the normal image reconstruction as a byproduct, thus still producing competative results where reconstruction dependent methods such as f-AnoGAN fail to work on. The current version of our ASC-Net aims to solve the two-cut problem, which will be tasked to handle more than two selective cuts in the future. Theoretical understanding of the proposed network is also required, which is left as a future work.

\begin{figure}[h]
\centering
  \includegraphics[width=0.35\textwidth]{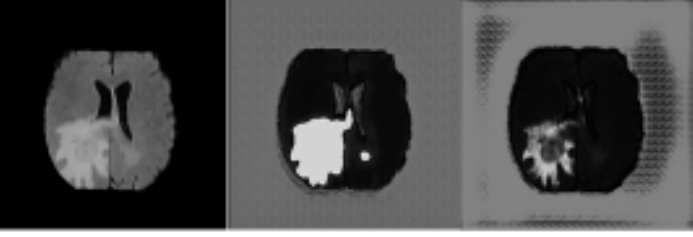}\\
  \includegraphics[width=0.35\textwidth]{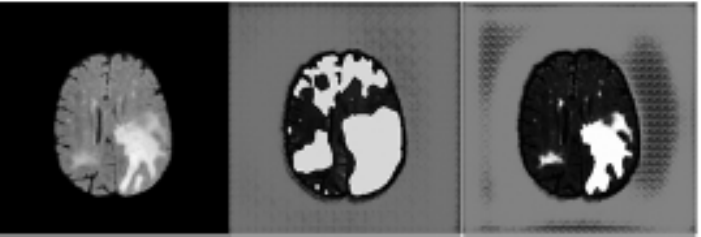}
\caption{Termination of network training affects the reconstruction result. Left to right columns in each image: the input images, the images reconstructed via two cycles in the first stage and one in the second stage, and the images reconstructed via adding one cycle in the second stage.}
\label{fig:train_more}
\end{figure}

\begin{figure*}[!ht]
\centering
  \includegraphics[width=0.117\textwidth]{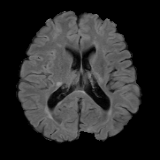}
  \includegraphics[width=0.117\textwidth]{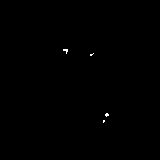}
  \includegraphics[width=0.117\textwidth]{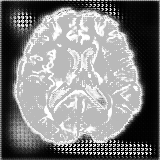}
  \includegraphics[width=0.117\textwidth]{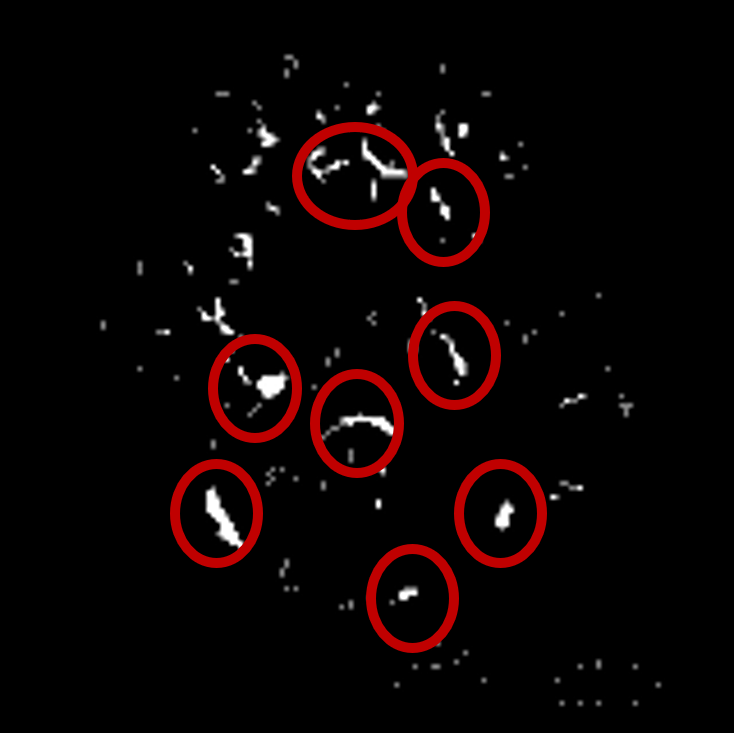} \;
  \includegraphics[width=0.117\textwidth]{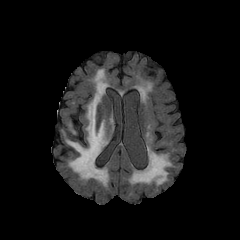}
  \includegraphics[width=0.117\textwidth]{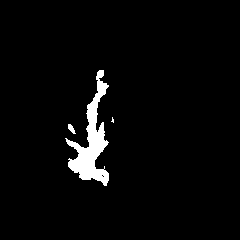}
  \includegraphics[width=0.117\textwidth]{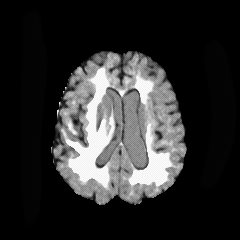}
  \includegraphics[width=0.117\textwidth]{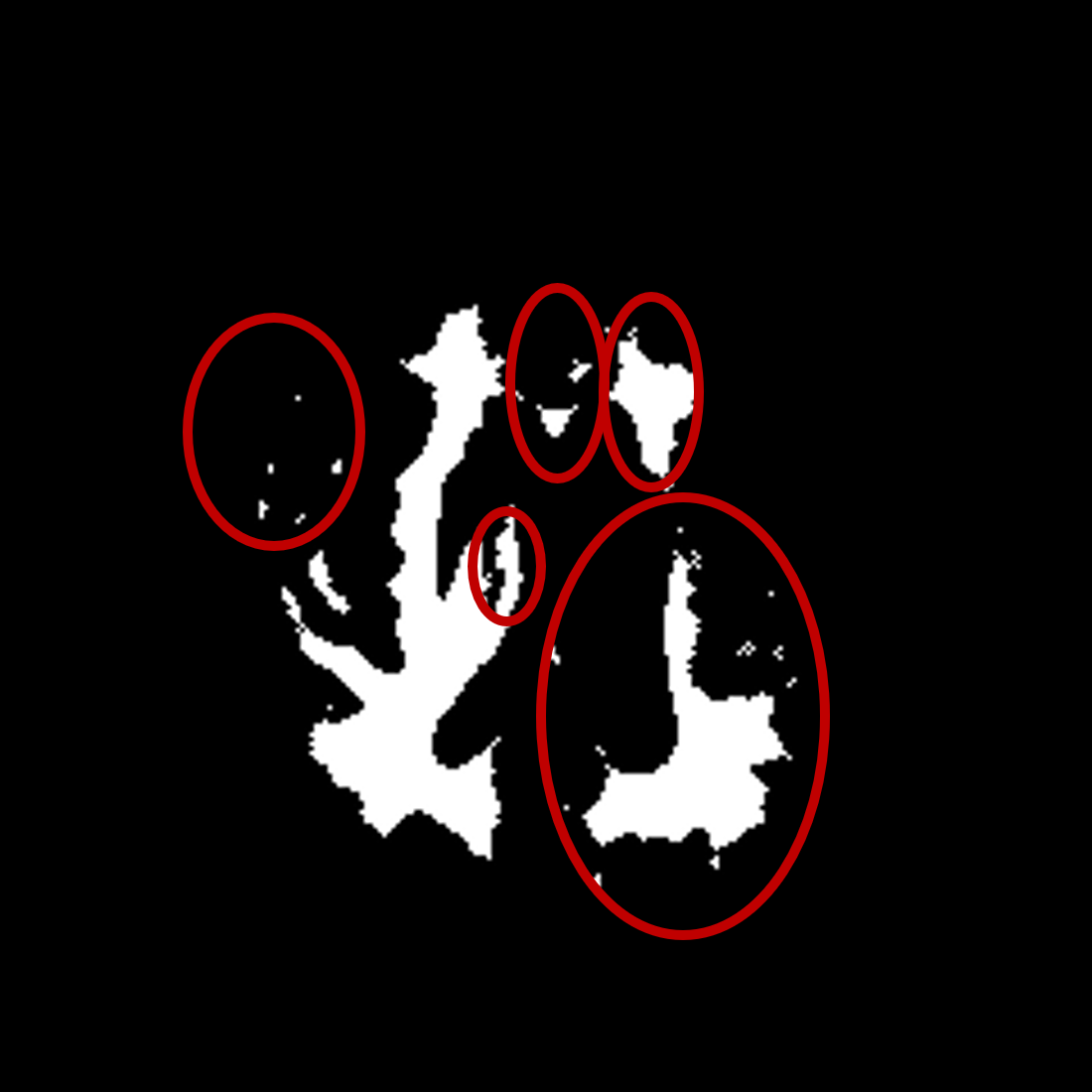}
 \caption{Demonstrating potential better interpretation on prediction compared to the ground truth, with samples collected from MS-SEG (left) and BraTS (right). Left to right: the input image, the corresponding ground-truth mask, our reconstructed image, our prediction with possible missing lesions highlighted in red ellipses.}
\label{fig:additional_anomalies}
\end{figure*}

\vspace{0.05in}
\noindent
\textbf{Termination and stability.} The termination point of this network training is periodic. The general guideline is that the peaks should be well separated and we terminate our algorithm at three or four peak separation. However, continuing to train further may not always result in the improvement for the purpose of segmentation due to accumulation of holes as shown in  Fig.~\ref{fig:train_more}, even though visually the anomaly is captured in more intricate detail. However, we encourage training longer as it reduces false positives and provides detailed anomaly reconstruction, though the Dice metric might not account for it. In our experiments, we specify the number of cycles in each stage. However, due to the random nature of the algorithm and the lack of a particular purpose and guidance, the peak separation may occur much earlier, then training should be stopped accordingly. The reported network in our Brats-2019 experiments has an average Dice score of $6\%$ over the network trained longer as shown in Fig.~\ref{fig:train_more}. Regarding the stability, Figure~\ref{fig:stability} demonstrates an anomaly estimated by different networks that are trained with different number of training cycles. We observe that while the appearance of $I_{ro}$ changes, we still obtain the anomaly as a separate cut since our framework works without depending on the quality of reconstruction.


\begin{figure*}[!h]
\centering

   \includegraphics[width=0.195\textwidth]{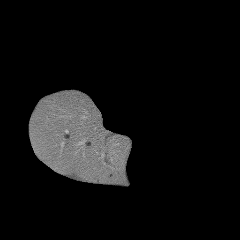}
  \includegraphics[width=0.195\textwidth]{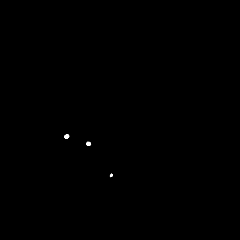}
  \includegraphics[width=0.195\textwidth]{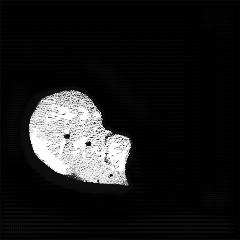}
  \includegraphics[width=0.195\textwidth]{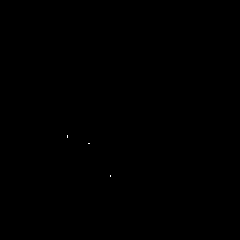}
  \includegraphics[width=0.195\textwidth]{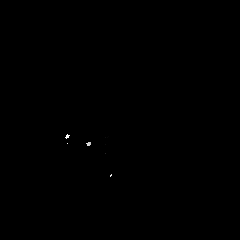}

\caption{Thresholding makes a difference, especially for small lesions or tumors. Left to right: the input image sampled from the LiTS dataset, the corresponding ground-truth mask, our reconstructed image, the mask generated with a threshold at 242 (used for the entire dataset), and the mask generated with a threshold at 238 (selected based on the reconstructed image of this subject).}
\label{fig:lits_err}
\end{figure*}


\vspace{0.05in}
\noindent
\textbf{Limitations.\label{limitations}} The low Dice scores reported for the public datasets could be because we have to select {\it non-tumor} slices as our reference distribution, which does not account for other co-morbidities. This affects the performance of the framework as it has no other guidance and would consider co-morbidities as detected anomaly as well. This statement is partially supported by our better brain tumor segmentation results on the private dataset, which has better {\it healthy} scans for forming the reference distribution. Moreover, although having more false positives due to the potential co-morbidities, our method provides possibility of bringing other potential anomalies into the user's attention or even better anomaly masks than the ground truth. As shown in Fig.~\ref{fig:Flair_vs_t2_1} and~\ref{fig:additional_anomalies}, compared to the ground-truth tumor masks, our method identifies more regions that present similar appearance as annotated tumor regions.   

One possible improvement on our method is the automatic or subject-specific selection of the threshold, which is used to obtain the final segmentation mask. In the current work, we choose one threshold for the entire test set, which is probably not the optimal solution. For example, in the liver lesion segmentation experiment, if we were to take a threshold at 238 based on the single sample's rightmost peak, we get a better final mask for the particular sample, compared to taking the threshold at 242 based on the entire data set (see Fig.~\ref{fig:lits_err}).

Regarding the use of our post-processing scheme for the public datasets, we observe that it could remove some noise like the faint liver boundary as shown in Fig.~\ref{fig:result_brats}; however, at the same time it could remove the small tiny lesions detected by the network, as shown in Fig.~\ref{fig:lits_err}. Improvement over the network output without post-processing needs further investigation, which is left for the future work.


Overall, our framework demonstrates critical insights that may be missed from annotations across datasets curated from different institutions;
it can also locate very tiny small tumors accurately if an appropriate threshold is taken.
A better use of our method could be to assist in the initial discovery of anomalous or novel markers. Following the initial discovery, human inspection or domain-knowledge-guided post-processing could provide better segmentation results to be used in practice. 


\section*{Acknowledgements}
This work was supported by Shanghai Municipal Science and Technology
Major Project 2021SHZDZX0102 and NSF 1755970. 

{\small
\bibliographystyle{plain}
\bibliography{refs}
}
\end{document}